\documentclass[aps,amssymb,amsmath,pra,reprint,noshowpacs,superscriptaddress]{revtex4-1}
\usepackage[utf8]{inputenc}
\usepackage{amsthm}
\usepackage{mathtools}
\usepackage{physics}
\usepackage{xcolor}
\usepackage{graphicx}
\usepackage[left=23mm,right=13mm,top=35mm,columnsep=15pt]{geometry} 
\usepackage{adjustbox}
\usepackage{placeins}
\usepackage[T1]{fontenc}
\usepackage{lipsum}
\usepackage{csquotes}
\usepackage{siunitx}
\usepackage[english]{babel}

\usepackage{helvet}
\usepackage{scalefnt} 
\usepackage{natbib}
\usepackage{hyperref}
\usepackage{appendix}
\usepackage{color}
\usepackage{comment}
\hypersetup{
	colorlinks   = true, 
	urlcolor     = blue, 
	linkcolor    = blue,
	citecolor   = blue 
}
\begin{document}

\scalefont{1.0}

\title{Scalable all-optical cold damping of levitated nanoparticles}

\author{Jayadev~\surname{Vijayan}}
\email[Correspondence email address: ]{jvijayan@ethz.ch}
\author{Zhao~\surname{Zhang}}
\affiliation{Photonics Laboratory, ETH Z{\"u}rich, 8093 Z\"urich, Switzerland}
\author{Johannes~\surname{Piotrowski}}
\affiliation{Photonics Laboratory, ETH Z{\"u}rich, 8093 Z\"urich, Switzerland}
\author{Dominik~\surname{Windey}}
\affiliation{Photonics Laboratory, ETH Z{\"u}rich, 8093 Z\"urich, Switzerland}
\author{Fons~\surname{van der Laan}}
\affiliation{Photonics Laboratory, ETH Z{\"u}rich, 8093 Z\"urich, Switzerland}
\author{Martin~\surname{Frimmer}}
\affiliation{Photonics Laboratory, ETH Z{\"u}rich, 8093 Z\"urich, Switzerland}
\author{Lukas~\surname{Novotny}}
\affiliation{Photonics Laboratory, ETH Z{\"u}rich, 8093 Z\"urich, Switzerland}

\date{\today} 

\begin{abstract}

The field of levitodynamics has made significant progress towards controlling and studying the motion of a levitated nanoparticle.
Motional control relies on either autonomous feedback via a cavity or measurement-based feedback via external forces. 
Recent demonstrations of measurement-based ground-state cooling of a single nanoparticle employ linear velocity feedback, also called cold damping, and require the use of electrostatic forces on charged particles via external electrodes.
Here we introduce a novel all-optical cold damping scheme based on spatial modulation of the trap position that is scalable to multiple particles.
The scheme relies on using programmable optical tweezers to provide full independent control over trap frequency and position of each tweezer.
We show that the technique cools the center-of-mass motion of particles down to $17\,$mK at a pressure of $2 \times 10^{-6}\,$mbar and demonstrate its scalability by simultaneously cooling the motion of two particles.
Our work paves the way towards studying quantum interactions between particles, achieving 3D quantum control of particle motion without cavity-based cooling, electrodes or charged particles, and probing multipartite entanglement in levitated optomechanical systems.

\end{abstract}

\maketitle

\label{sec:introduction}
\centerline{\textbf{Introduction}}
\vspace{1mm}

The success of quantum theory has led to the emergence of experimental platforms seeking to harness quantum properties to understand fundamental aspects of nature as well as to develop next generation technology.
Levitodynamics approaches this task by using an optical tweezer to suspend a dielectric nanoparticle in ultra-high vacuum, resulting in a highly isolated mechanical oscillator~\cite{Millen2020,Gonzalez-Ballestero2021}.
The level of control achieved in levitated optomechanical systems has led to cooling the center-of-mass (c.m.) motion of a nanoparticle to its quantum ground state~\cite{Delic2020,Tebbenjohanns2021,Magrini2021}, and the development of inertial~\cite{Monteiro2017,Timberlake2019,Monteiro2020,Ahn2020,Fons2021} and field sensors~\cite{Gambhir2016,Hempston2017,Hebestreit2018}. 

Scaling up to multiple particles will open up new avenues of research, including probing quantum correlations and entanglement~\cite{Chauhan2020,Brandao2021,Kotler2021,Lepinay2021}, complex phases emerging from interacting particles~\cite{Reimann2015,Landig2016,Bernien2017,Liu2020,Periwal2021} and sensing of weak forces ~\cite{Bressi2002,Wang2021,Quinn2001,Li2018}.
Investigating these phenomena with levitated optomechanical systems requires three ingredients: 1) trapping of multiple particles, 2) cooling their motional degrees of freedom, and 3) controlling the interactions between them.
Here we introduce a scalable experimental platform capable of optically trapping multiple dielectric nanoparticles with tunable separation, based on techniques developed by the cold atom community~\cite{Kaufman2014,Endres2016,Daniel2016,Ebadi2021}.
The particles are trapped using a programmable array of optical tweezers generated by radiofrequency (rf) tones driving an acousto-optical deflector (AOD), allowing independent control of both the trap position and trap frequency of each particle.
We then introduce a novel, all-optical linear feedback cooling scheme that cools the motional degrees of freedom of each particle along the AOD axis by modulating the spatial position of the optical traps at the trap frequency, with the optical intensity gradient providing the restoring force.
We demonstrate the scalability of this cold damping scheme by simultaneously cooling the c.m.\ motion of a pair of particles.

Our scalable cooling scheme also stands out by its simplicity~\cite{Dania2021,Bang2020}.
It does not rely on optical cavities~\cite{Delic2019,Windey2019,Meyer2019,Delic2020}, near-field probes~\cite{Wilson2015}, charged particles or external electrodes~\cite{Tebbenjohanns2019colddamping,Tebbenjohanns2021}. 
In fact, the use of electrostatic forces severely limits the potential use of levitated optomechanical systems as sensing devices or platforms to probe quantum entanglement~\cite{Rudolph2020}, making our all-optical scheme uniquely suited for such applications.
The all-optical cold damping scheme thus serves as a complementary technique to parametric cooling~\cite{Gieseler2012,Jain2016} and as an alternative to electrostatic linear feedback cooling for multiple particles independent of their charge.
By utilizing the same tweezer for trapping, detection and cooling, we simplify the components required for cold damping in levitodynamics experiments.

\vspace{5mm}
\label{sec:setup}
\centerline{\textbf{Experimental setup}}
\vspace{1mm}

\begin{figure*}
    \centering
    \includegraphics[width = 15cm]{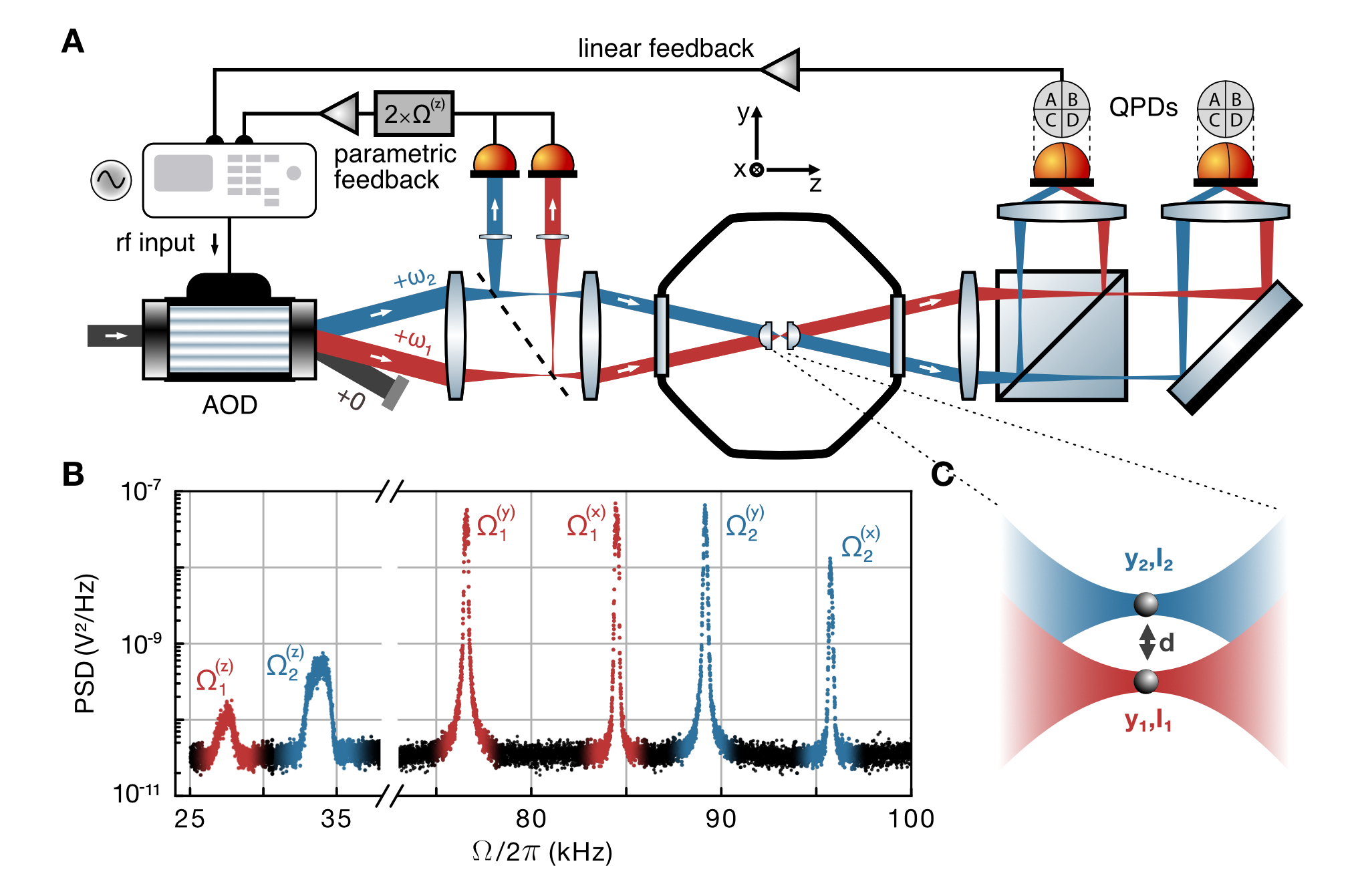}
    \caption{\textbf{Programmable tweezers for multi-particle trapping.} \textbf{A}.~Schematic of the experimental system.
    An array of beams with different optical frequencies $\omega_i$ is generated by sending multiple radio-frequency (rf) tones to an acousto-optical deflector (AOD). 
    Inside a vacuum chamber, the beams are focused by an aspheric lens to form a tweezer array along the $y$ direction to trap nanoparticles.
    The motion along the three spatial directions of each particle is detected by measuring the light scattered by the particles on photodetectors in homodyne schemes. 
    Both the forward- and backward-scattered light detectors are used to perform active feedback cooling of the particle motion (see text).
    An additional quadrant photodiode (QPD) in the forward-scattered direction is used as an out-of-loop detector.
    \textbf{B}.~The power spectral density (PSD) of the signal detected by the out-of-loop QPD. 
    The colored peaks correspond to the c.m.\ motion of both particles along $x$, $y$ and $z$ directions at a pressure of $p_\mathrm{gas} = 1 \times 10^{-4}\,$ mbar.
    \textbf{C}.~The setup provides independent control over the position $y_i$ and trap frequency (via tweezer intensity $I_i$) of each particle in the array by programming the frequency and amplitude of the rf tones being fed into the AOD.
    The spatial separation $d$ between the two tweezers is tunable up to $20~\mu$m.}
    \label{fig:setup}
\end{figure*}

Our mechanical oscillators are single spherical SiO$_2$ nanoparticles of nominal diameter $177\,$nm, which are levitated in high vacuum using optical tweezers at wavelength $\lambda = 1550\,$nm and optical power $P = 200\,$mW, focused by a high numerical aperture (NA $ = 0.75$) lens. 
The setup makes use of an acousto-optical deflector (AOD) to generate a tweezer array~\cite{Endres2016} and a telescope to angle the beams towards the high-NA lens, resulting in optical traps for multiple nanoparticles, as shown in Fig.~\ref{fig:setup}A.
The tweezer array is generated by sending multiple radiofrequency (rf) tones {$\omega_i$} as input to the AOD, resulting in several first-order deflected beams propagating along the $z$ axis, shifted in frequency by $\omega_{\text{i}}$ and emerging with different geometric angles along the $y$ axis.
By changing the frequencies of the rf tones, we steer the first-order beams and as a result, change the positions of the optical tweezers.
By construction, the tweezers are at different optical frequencies, which highly suppresses dipole-dipole interactions between the particles~\cite{Rieser2022}.
In our experiment, the spatial separation between the particles can be tuned up to $d = 20$ $\mu$m, limited by the geometry of the setup and the bandwidth of the deflector.
In addition, we control the amount of optical power in the diffracted beams by tuning the amplitude of the rf tones sent to the AOD.
Tuning the optical power changes the stiffness of the trap, which can be used to parametrically cool the particle motion~\cite{Gieseler2012}.


\begin{figure*}
    \centering
    \includegraphics[width = 12.9cm]{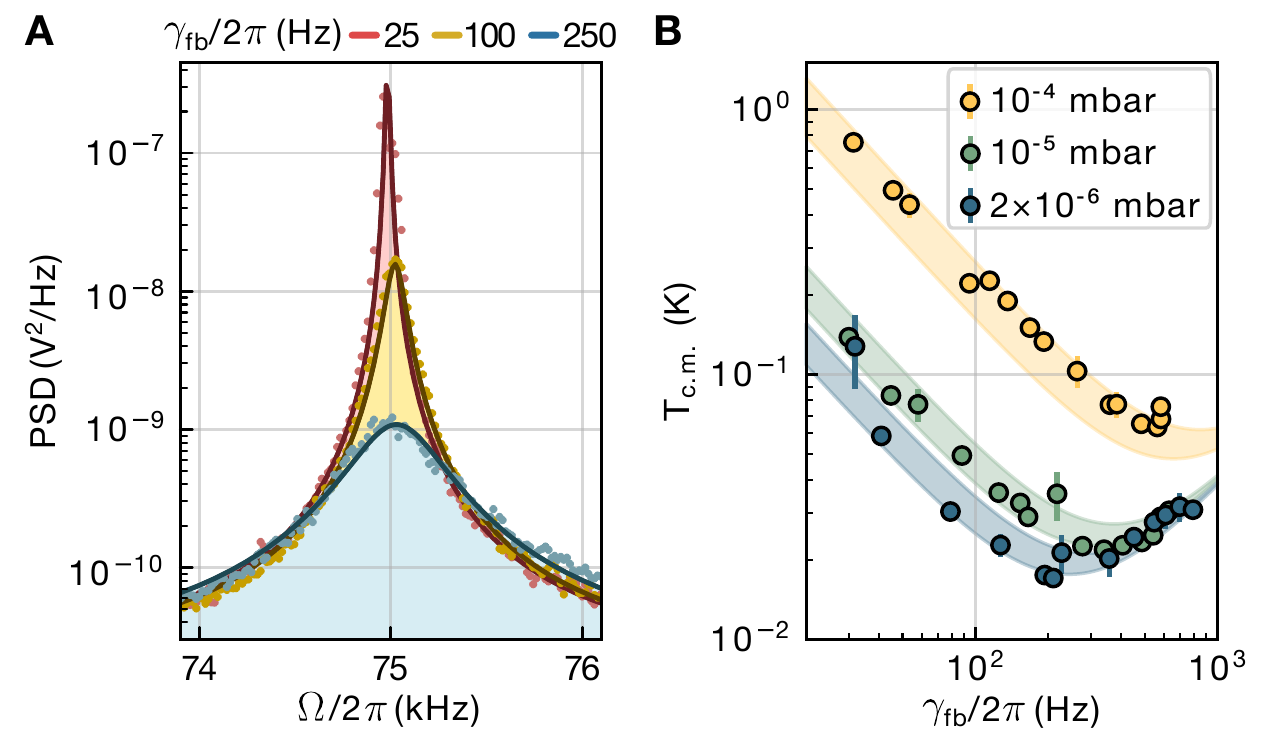}
    \caption{\textbf{All-optical cold damping by modulation of spatial position of the tweezer.}
    \textbf{A}.~Measurement of the particle motion along the $y$ axis, using the out-of-loop detector.
    PSDs of the motional peak are shown at the pressure of $p_\mathrm{gas} = 1 \times 10^{-4}\,$ mbar at low (red), medium (yellow) and high (blue) feedback gains.
    Solid lines are Lorentzian fits to the data.
    \textbf{B}.~Measured center-of-mass temperatures (T$_\text{c.m.}$) at three different pressures as a function of the feedback damping rate $\gamma_{\text{fb}}$.
    At higher gains, when the PSD of the particle motion is cooled close to the noise floor, the feedback signal becomes dominated by noise and the temperatures rise again (see text).
    At the lowest pressure and optimal gain, the particle motion is cooled to a temperature of $17\,$mK.
    Shaded area corresponds to estimated temperature ranges given fluctuations in pressure during data-collection, as measured from reheating experiments.
    }
    \label{fig:linearcooling}
\end{figure*}

The light scattered by the particles in the forward direction, as well as light passing by the particles are interfered on two quadrant photodetectors (QPD), resulting in a homodyne detection of the particles' c.m.\ motion \cite{Gieseler2012}. 
By subtracting the left (upper) and right (lower) quadrant from each other we obtain a detection signal that is primarily sensitive to the $y$ ($x$) motion of the particles (see Fig.~\ref{fig:setup}B). 
Due to the tilted alignment of the beam we also detect the $z$ motion on the QPDs.
The two QPDs act as in and out-of-loop detectors to linearly cool and independently detect the particles' $x$ and $y$ motion \cite{Tebbenjohanns2019colddamping,Dania2021} (see Supplementary).
The light scattered by each particle in the backward direction is split off and interfered on detectors with backreflections from the vacuum window, detecting primarily the $z$ motion of the particles~\cite{Tebbenjohanns2019PRA}. 
For the experiments that follow, we pre-cool the particle motion along the $z$-axis using parametric feedback to avoid nonlinearities in the trapping potential~\cite{Gieseler2012}.
To derive particle c.m.\ temperatures we rely on the calibration techniques outlined in \cite{Hebestreit2018Calibration}.


\vspace{5mm}
\label{sec:cooling}
\centerline{\textbf{All-optical feedback cooling of a single particle}}
\vspace{1mm}

Active, measurement-based feedback cooling schemes have been deployed to cool the c.m.\ motion of single nanoparticles in several recent experiments, and can generally be classified into two types - parametric and linear feedback cooling.
In parametric feedback cooling, the measured signal is filtered, phase-shifted and fed back at twice the frequency of particle motion.
Typically, parametric feedback is applied by modulating the optical power of the tweezer~\cite{Gieseler2012}, and has been used to cool the c.m.\ motion to several tens of phonons~\cite{Jain2016}.
In linear feedback cooling, or cold damping, the measured signal is filtered, phase-shifted and applied as a direct force at the trap frequency of the particle.
Cold damping schemes have been used to cool the motion of charged particles \cite{Steixner2005,Bushev2006,Iwasaki2018,Tebbenjohanns2019colddamping} by applying a viscous electrostatic force at the trap frequency.
Such schemes have recently been used to cool the motion of a charged nanoparticle along the beam propagation direction ($z$) to the quantum ground state~\cite{Tebbenjohanns2021,Magrini2021}.

Our all-optical linear feedback does not require electrodes or charged particles.
Rather than modulating the voltage applied to external electrodes, we modulate the spatial position of the tweezer along the AOD axis ($y$) at the frequency of the particle motion.
The gradient force acts as a restoring force, which increases with the relative distance between tweezer and particle.
In our experiment, trapping beams are generated by applying rf tones to the AOD. 
By appropriately choosing the amplitude and frequency of the tones, both parametric and linear feedback cooling can be performed simultaneously.
Amplitude modulation of the rf tone at twice the trap frequency parametrically cools the particle motion whereas frequency modulation of the tone at the trap frequency linearly cools the particle by spatial displacement. 

Similar to other cold damping schemes \cite{Cohadon1999,Poggio2007,Wilson2015,Rossi2018,Tebbenjohanns2019colddamping}, the feedback circuit itself consists of a digital filter that electronically processes the in-loop QPD signal in real-time. 
The filter comprises of two components - a delay line to shift the phase of the frequencies near $\Omega^{(y)}$ by $\pi/2$ and a variable amplifier which sets the gain of the signal. 
The filtered signal is sent to the frequency modulation input of the function generator driving the AOD.
A frequency modulation of $\Delta\omega_i$ around the central frequency $\omega_i$ results in a periodic spatial displacement of the tweezer, with the amplitude of displacement given by the gain set in the filter (see Supplementary).

In addition to using the in-loop QPD for feedback control, we use a second out-of-loop QPD as an independent measure of the particle motion.
We record the power spectral density (PSD) of the particle at different feedback gains and gas pressures to estimate the cooling performance (see Fig. \ref{fig:linearcooling}).
In the regime where gas damping is negligible compared to feedback damping ($\gamma_\text{gas} \ll \gamma_\text{fb}$) the c.m.\ temperature $T_\text{c.m.}$ can be modeled as \cite{Poggio2007,Tebbenjohanns2019colddamping}


\begin{align}
\label{eqn:Tcm_cold_damping}
    T_\text{c.m.} = \frac{\gamma_\text{gas} T_\text{gas}}{\gamma_\text{fb}} + \frac{\pi m \gamma_\text{fb} {\Omega^{(y)}}^2 S_\text{nn}}{k_\text{B}}.
\end{align}

Here $T_\text{gas}$ is the temperature of the surrounding gas molecules, $m$ and $\Omega^{(y)}$ are the nanoparticle mass and mechanical frequency of the feedback cooled axis, and $S_\text{nn}$ denotes the in-loop QPD noise floor.
The datapoints in Fig.~\ref{fig:linearcooling}B are obtained from the area under a calibrated power spectral density \cite{Hebestreit2018Calibration} whereas the shaded regions represent the model described by Eq.~\ref{eqn:Tcm_cold_damping}. The parameters of the model are not fitted but obtained from reheating experiments, described in the next section, and the calibrated noise floor of the in-loop detector. The width of the shaded region is given by pressure fluctuations during the experiments.
For small feedback gains $\gamma_{\text{fb}}$, the c.m.\ temperature of the particle decreases with increasing gains, in agreement with the first term of Eq.~\ref{eqn:Tcm_cold_damping}.
However, for large gains, when the signal approaches the noise floor, noise is fed back by the loop, heating the particle motion as predicted by the second term of Eq.~\ref{eqn:Tcm_cold_damping}.
At the optimal feedback gain, the c.m.\ temperature of the particle is cooled down to $17\,$mK.

\vspace{5mm}
\label{sec:noneq}
\centerline{\textbf{Non-equilibrium dynamics}}
\vspace{1mm}

\begin{figure*}
    \centering
    \includegraphics[width = 17.2cm]{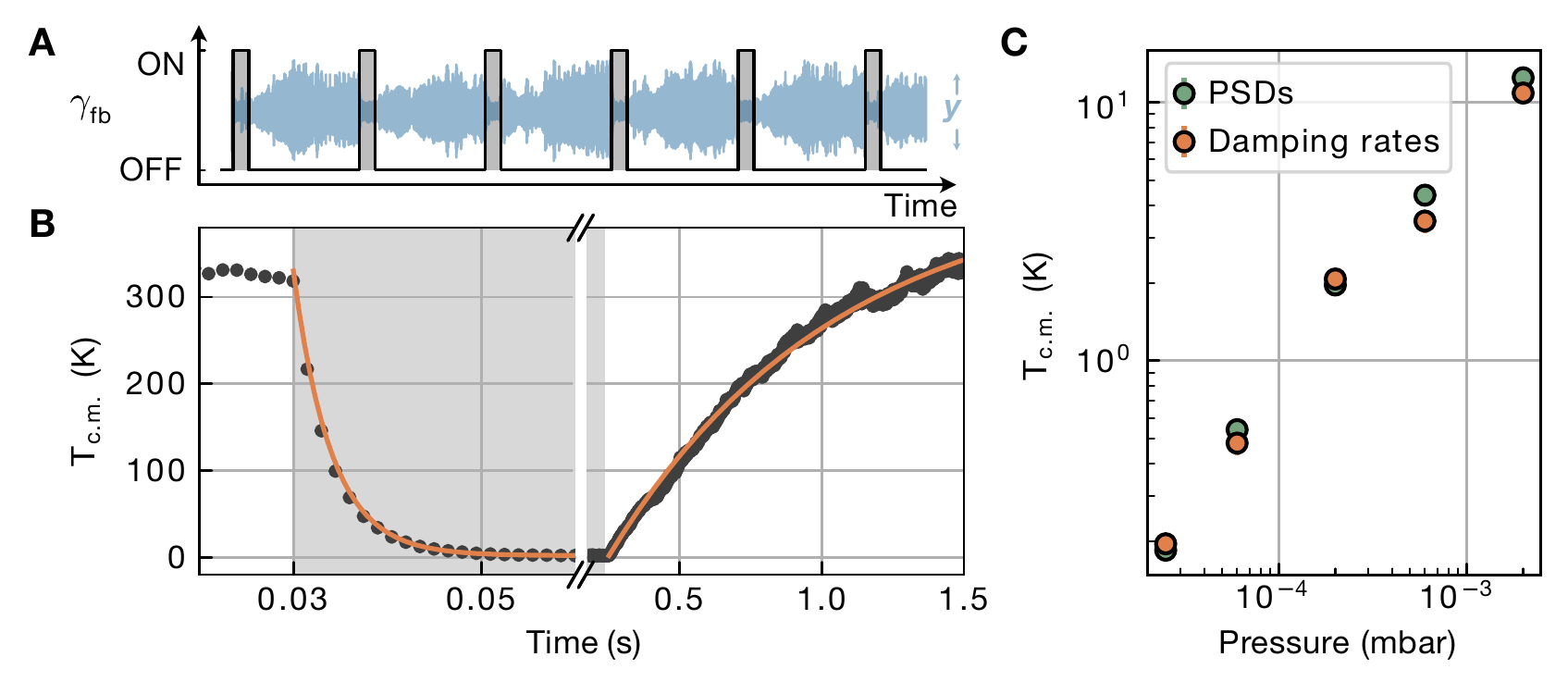}
    \caption{\textbf{Non-equilibrium dynamics of a feedback-cooled particle.}
    \textbf{A}.~Ring-down and reheating measurements are performed by periodically switching on ($200\,$ms) and off ($1300\,$ms) the cold damping feedback loop.
    A few example time traces of the non-equilibrium dynamics of the $y$ motion are shown in blue. 
    \textbf{B}.~We average $90$ reheating cycles to obtain the temperature of the particle as a function of time.
    The fits (orange) are used to extract the feedback damping and gas damping rates, shown here for measurements at a pressure of $p_\mathrm{gas} = 2 \times 10^{-4}\,$ mbar.
    Gray shaded regions indicate when the feedback is turned on.
    \textbf{C}.~The temperature extracted from the ratio of damping rates (orange) are compared with temperatures extracted from the area under the PSDs of long time traces (green) at a fixed low gain of $\gamma_{\text{fb}} = 2\pi\times 42\,$Hz at different pressures.
    }
    \label{fig:reheat}
\end{figure*}

By construction of our feedback scheme, the spatial modulation of the tweezer is at the same frequency as the particle motion. 
As a consequence, the signals measured by the QPDs contain contributions from both the particle motion and the tweezer modulation.
To ensure that the extracted temperatures are reliable, we perform non-equilibrium measurements to independently extract the temperature from the ratio of damping rates of the feedback cooling $\gamma_{\text{fb}}$ and gas reheating $\gamma_{\text{gas}} T_{\text{gas}}$.

In a first step, we perform a ring-down sequence where the linear feedback cooling is turned on for $200\,$ms, allowing the particle to exponentially equilibrate from a high c.m.\ temperature $T_\text{gas}$ to a low c.m.\ temperature $T_\text{cold}$ (see Fig. \ref{fig:reheat}) according to $T_\text{c.m.}(t) = T_\text{cold} + (T_\text{gas} - T_\text{cold}) \exp(-\gamma_\text{fb}t) $ \cite{Gieseler2014}.
Then the feedback is abruptly switched off, and the particle reheats through gas collisions until it equilibrates with the background gas temperature $T_\text{gas}$, following the exponential growth given by $T_\text{c.m.}(t) = T_\text{gas} + (T_\text{cold} - T_\text{gas}) \exp(-\gamma_\text{gas}t) \approx T_\text{cold} + \gamma_\text{gas} T_\text{gas} t$ for small $t$.
As the reheating dynamics are very slow ($1/\gamma_\text{gas} > 3\,$s) at pressures $p < 10^{-4}\,$mbar, we limit the reheating time and therefore fit a linear slope instead of the full exponential curve.
The protocol is repeated $90$ times to extract both a feedback damping rate from the ring-down sequence and a gas damping rate from the reheating sequence.
Crucially, measuring the particle motion during ring-down and reheating cycles allows us to track the PSDs of the particle continuously as it is cooled under the cold damping scheme and then reheated by the gas when the feedback is turned off, i.e, in the absence of tweezer modulation.

We repeat the non-equilibrium protocol and extract the damping rates for different pressures and fixed gain.
The feedback damping rate $\gamma_{\text{fb}}$ is found to be independent of pressure and the gas damping rate $\gamma_{\text{gas}} T_{\text{gas}}$ decreases linearly with pressure (see Supplementary). 
According to Eq.~\ref{eqn:Tcm_cold_damping}, in a regime of low feedback gain, we expect the particle c.m.\ temperature to be given by the ratio of gas reheating and feedback cooling rates $T_\text{c.m.} = \gamma_\text{gas} T_\text{gas} / \gamma_\text{fb}$. We compare the temperatures extracted from the ratio of damping rates at different pressures to the temperatures extracted from the area under the PSDs in Fig.~\ref{fig:reheat}C and find them in good agreement across the entire pressure range.

\vspace{5mm}

\label{sec:simultaneous}
\centerline{\textbf{Simultaneous two-particle cooling}}
\vspace{1mm}

\begin{figure*}
    \centering
    \includegraphics[width = 17.2cm]{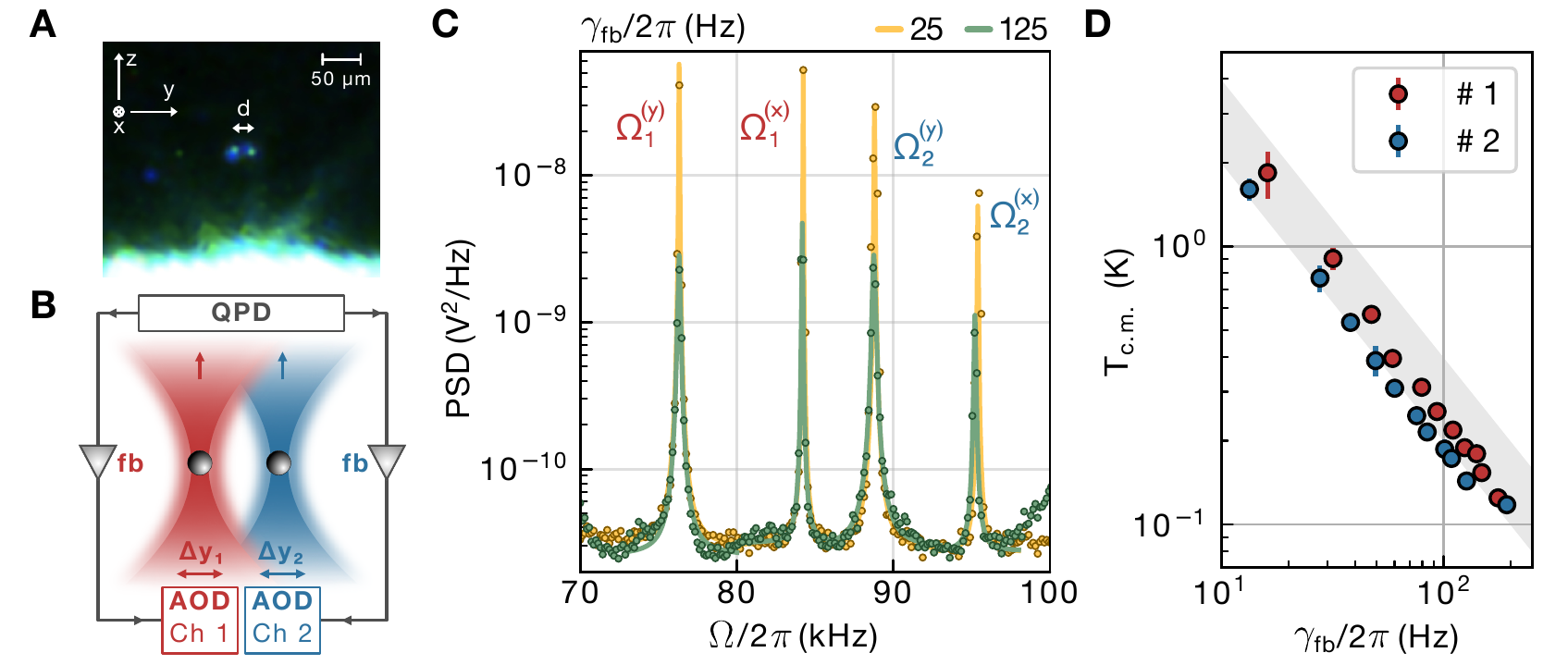}
    \caption{\textbf{Simultaneous all-optical cold damping of two particles.}
    \textbf{A}.~Image taken of two nanoparticles levitated simultaneously by a high NA lens in a vacuum chamber, with a tunable spatial separation $d$.
    For the experiments that follow, the separation is fixed at $d = 4.6~\mu$m.
    \textbf{B}.~Schematic of simultaneous all-optical cold damping of two particles.
    Signals of the motion of the two particles are measured by the in-loop QPD, spectrally separated, and fed back to different channels of a function generator driving the AOD. 
    The signal of each particle is spectrally separated from the other by having the tweezers at slightly different optical powers, resulting in different trap frequencies.
    \textbf{C}.~PSDs of the motion of two particles as measured by the in-loop QPD, at low (yellow) and high (green) feedback gain.
    Solid lines are Lorentzian fits to the data.
    \textbf{D}.~Simultaneous cooling of two particles at different gains.
    Increasing the gain on each feedback loop cools the center-of-mass motion of the corresponding particle.
    At a pressure of $p_\mathrm{gas} = 1 \times 10^{-4}\,$ mbar, both particles are cooled to a temperature close to $100\,$mK at the highest applied feedback gain.
    The shaded area corresponds to estimated temperature ranges given fluctuations in pressure during data-collection, as measured from reheating experiments.
    }
    \label{fig:twoparticle}
\end{figure*}

One of the benefits of linear feedback cooling via modulation of displacement is the scalability to multiple particles.
Since each rf tone entering the deflector is generated independently by different channels of a function generator and the optical power and spatial position of each tweezer are determined solely by the amplitude and frequency of the tone, a separate feedback loop can be applied to each tweezer (see Fig.~\ref{fig:twoparticle}B).
We split the output signal of the $y$-channel of the QPD into two paths, each going into separate electronic filters.
In both paths, notch filters are used to filter out all frequencies except the motional frequency of interest $\Omega_{i}^{(y)}$. 
Appropriate delays and gains are added to each path, before the signal is fed back as frequency modulation input to the function generators.
The setup can cool each particle independently and with adjustable cooling rates. 

Combining the linear feedback scheme with parametric feedback via amplitude modulation at twice the $z$ frequencies of particle motion $\Omega_i^{(z)}$ enables simultaneous two-axis cooling of both particles.
The PSD of two particles when cold damping is applied simultaneously to both particles at a pressure of $p_\mathrm{gas} = 1 \times 10^{-4}\,$ mbar, is shown in Fig.~\ref{fig:twoparticle}C.
Due to coupling between the different axes of the particle’s oscillatory motion arising from residual non-linearities in the potential, cooling the $y$ and $z$ motion also cools the $x$ degree of freedom.
However, this effect is expected to vanish as the particle is cooled further and its motional amplitudes are reduced.
Small differences in particle size or detector alignment of the tweezers can lead to a slight asymmetry in the detection efficiency of the two particles' motion.
Consequently, we choose slightly different gain settings for the linear filter of the feedback loop, to obtain the same cooling rates $\gamma_{\text{fb}}$ for both particles (see Supplementary). 
With these adjustments, both particles can be simultaneously cooled to the same temperature, as shown in Fig. \ref{fig:twoparticle}D.

With the current detection scheme, a single QPD in the forward scattering direction is sufficient to detect and feedback-cool several particles as long as their motional peaks are spectrally separated by $5\,$kHz, limited by the filter bandwidths in the feedback loop.
A straightforward way to scale up the system is to use a laser source with higher power, allowing us to spread the trap frequencies over a larger range.
Our cold damping scheme can also be used to perform feedback cooling of particles with degenerate trap frequencies, by using a heterodyne detection scheme (see Supplementary) or by collecting the light scattered by each particle on separate QPDs. 


\vspace{5mm}
\label{sec:conclusions}
\centerline{\textbf{Conclusion and outlook}}
\vspace{1mm}

We have demonstrated an experimental platform capable of trapping and cooling multiple nanoparticles.
By programming the rf tones driving the AOD, we achieve independent control of both the position and trap frequency of each particle.
We introduced an all-optical cold damping scheme based on modulation of spatial displacement, which allows the usage of a single tweezer to perform trapping, cooling and detection of the particle motion.
Finally, we showed that the feedback cooling techniques are scalable to two particles.

The lowest temperatures we currently achieve are limited both by the background pressure in the chamber and the detection efficiency of the particle motion.
Lower pressures can be achieved by baking the chamber, and will result in lower temperatures until we become limited by radiation pressure shot noise.
Adding lenses or an optical cavity along the $y$ axis will provide a higher detection efficiency for the light scattered by the particle along the transverse direction~\cite{Tebbenjohanns2019PRA}.
An optical cavity can be used to enable tunable long-range interactions between spatially separated nanoparticles~\cite{Periwal2021,Debnath2016}, to probe hybrid modes~\cite{Reimann2015,delosRiosSommer2021,Toros2021}, and to potentially generate motional quantum entanglement~\cite{Rudolph2020,Rudolph2022}.
Furthermore, the all-optical cold damping scheme can be readily extended to both transverse directions ($x$ and $y$) by introducing a second AOD~\cite{Ebadi2021}, oriented orthogonal to the first.
In this way the motion of multiple particles can be cooled along both $x$ and $y$ axes.
Combined with parametric cooling along $z$, our protocol can achieve 3D cooling of multiple levitated particles by entirely optical means.

\emph{Note:} We recently became aware of related work on using optical lattices to cool the motion of a single nanoparticle~\cite{Kamba2022}. 

This research was supported by the Swiss National Science Foundation (SNF) through the NCCR-QSIT programme (grant no. 51NF40-160591), European Union’s Horizon 2020 research and innovation programme under grant numbers 863132 (iQLev) and 951234 (Q-Xtreme), and ETH Grant ETH-47~20-2.
We thank L. Bobzien, E. Bonvin, M. Cavigelli, U. Delic, J. Gao, M. L. Mattana, A. Militaru, M. Rossi, F. Tebbenjohanns, N. C. Zambon and J. Zielinska for valuable input and discussions.

\bibliographystyle{apsrev4-1}
\bibliography{bibliography_main}

\begin{thebibliography}{55}%
\makeatletter
\providecommand \@ifxundefined [1]{%
 \@ifx{#1\undefined}
}%
\providecommand \@ifnum [1]{%
 \ifnum #1\expandafter \@firstoftwo
 \else \expandafter \@secondoftwo
 \fi
}%
\providecommand \@ifx [1]{%
 \ifx #1\expandafter \@firstoftwo
 \else \expandafter \@secondoftwo
 \fi
}%
\providecommand \natexlab [1]{#1}%
\providecommand \enquote  [1]{``#1''}%
\providecommand \bibnamefont  [1]{#1}%
\providecommand \bibfnamefont [1]{#1}%
\providecommand \citenamefont [1]{#1}%
\providecommand \href@noop [0]{\@secondoftwo}%
\providecommand \href [0]{\begingroup \@sanitize@url \@href}%
\providecommand \@href[1]{\@@startlink{#1}\@@href}%
\providecommand \@@href[1]{\endgroup#1\@@endlink}%
\providecommand \@sanitize@url [0]{\catcode `\\12\catcode `\$12\catcode
  `\&12\catcode `\#12\catcode `\^12\catcode `\_12\catcode `\%12\relax}%
\providecommand \@@startlink[1]{}%
\providecommand \@@endlink[0]{}%
\providecommand \url  [0]{\begingroup\@sanitize@url \@url }%
\providecommand \@url [1]{\endgroup\@href {#1}{\urlprefix }}%
\providecommand \urlprefix  [0]{URL }%
\providecommand \Eprint [0]{\href }%
\providecommand \doibase [0]{http://dx.doi.org/}%
\providecommand \selectlanguage [0]{\@gobble}%
\providecommand \bibinfo  [0]{\@secondoftwo}%
\providecommand \bibfield  [0]{\@secondoftwo}%
\providecommand \translation [1]{[#1]}%
\providecommand \BibitemOpen [0]{}%
\providecommand \bibitemStop [0]{}%
\providecommand \bibitemNoStop [0]{.\EOS\space}%
\providecommand \EOS [0]{\spacefactor3000\relax}%
\providecommand \BibitemShut  [1]{\csname bibitem#1\endcsname}%
\let\auto@bib@innerbib\@empty
\bibitem [{\citenamefont {Millen}\ \emph {et~al.}(2020)\citenamefont {Millen},
  \citenamefont {Monteiro}, \citenamefont {Pettit},\ and\ \citenamefont
  {Vamivakas}}]{Millen2020}%
  \BibitemOpen
  \bibfield  {author} {\bibinfo {author} {\bibfnamefont {J.}~\bibnamefont
  {Millen}}, \bibinfo {author} {\bibfnamefont {T.~S.}\ \bibnamefont
  {Monteiro}}, \bibinfo {author} {\bibfnamefont {R.}~\bibnamefont {Pettit}}, \
  and\ \bibinfo {author} {\bibfnamefont {A.~N.}\ \bibnamefont {Vamivakas}},\
  }\href {\doibase 10.1088/1361-6633/ab6100} {\bibfield  {journal} {\bibinfo
  {journal} {Reports on Progress in Physics}\ }\textbf {\bibinfo {volume}
  {83}},\ \bibinfo {pages} {026401} (\bibinfo {year} {2020})}\BibitemShut
  {NoStop}%
\bibitem [{\citenamefont {Gonzalez-Ballestero}\ \emph
  {et~al.}(2021)\citenamefont {Gonzalez-Ballestero}, \citenamefont
  {Aspelmeyer}, \citenamefont {Novotny}, \citenamefont {Quidant},\ and\
  \citenamefont {Romero-Isart}}]{Gonzalez-Ballestero2021}%
  \BibitemOpen
  \bibfield  {author} {\bibinfo {author} {\bibfnamefont {C.}~\bibnamefont
  {Gonzalez-Ballestero}}, \bibinfo {author} {\bibfnamefont {M.}~\bibnamefont
  {Aspelmeyer}}, \bibinfo {author} {\bibfnamefont {L.}~\bibnamefont {Novotny}},
  \bibinfo {author} {\bibfnamefont {R.}~\bibnamefont {Quidant}}, \ and\
  \bibinfo {author} {\bibfnamefont {O.}~\bibnamefont {Romero-Isart}},\ }\href
  {\doibase 10.1126/science.abg3027} {\bibfield  {journal} {\bibinfo  {journal}
  {Science}\ }\textbf {\bibinfo {volume} {374}},\ \bibinfo {pages} {eabg3027}
  (\bibinfo {year} {2021})}\BibitemShut {NoStop}%
\bibitem [{\citenamefont {Deli{\'c}}\ \emph {et~al.}(2020)\citenamefont
  {Deli{\'c}}, \citenamefont {Reisenbauer}, \citenamefont {Dare}, \citenamefont
  {Grass}, \citenamefont {Vuleti{\'c}}, \citenamefont {Kiesel},\ and\
  \citenamefont {Aspelmeyer}}]{Delic2020}%
  \BibitemOpen
  \bibfield  {author} {\bibinfo {author} {\bibfnamefont {U.}~\bibnamefont
  {Deli{\'c}}}, \bibinfo {author} {\bibfnamefont {M.}~\bibnamefont
  {Reisenbauer}}, \bibinfo {author} {\bibfnamefont {K.}~\bibnamefont {Dare}},
  \bibinfo {author} {\bibfnamefont {D.}~\bibnamefont {Grass}}, \bibinfo
  {author} {\bibfnamefont {V.}~\bibnamefont {Vuleti{\'c}}}, \bibinfo {author}
  {\bibfnamefont {N.}~\bibnamefont {Kiesel}}, \ and\ \bibinfo {author}
  {\bibfnamefont {M.}~\bibnamefont {Aspelmeyer}},\ }\href {\doibase
  10.1126/science.aba3993} {\bibfield  {journal} {\bibinfo  {journal}
  {Science}\ }\textbf {\bibinfo {volume} {367}},\ \bibinfo {pages} {892}
  (\bibinfo {year} {2020})}\BibitemShut {NoStop}%
\bibitem [{\citenamefont {Tebbenjohanns}\ \emph {et~al.}(2021)\citenamefont
  {Tebbenjohanns}, \citenamefont {Mattana}, \citenamefont {Rossi},
  \citenamefont {Frimmer},\ and\ \citenamefont {Novotny}}]{Tebbenjohanns2021}%
  \BibitemOpen
  \bibfield  {author} {\bibinfo {author} {\bibfnamefont {F.}~\bibnamefont
  {Tebbenjohanns}}, \bibinfo {author} {\bibfnamefont {M.~L.}\ \bibnamefont
  {Mattana}}, \bibinfo {author} {\bibfnamefont {M.}~\bibnamefont {Rossi}},
  \bibinfo {author} {\bibfnamefont {M.}~\bibnamefont {Frimmer}}, \ and\
  \bibinfo {author} {\bibfnamefont {L.}~\bibnamefont {Novotny}},\ }\href
  {\doibase 10.1038/s41586-021-03617-w} {\bibfield  {journal} {\bibinfo
  {journal} {Nature}\ }\textbf {\bibinfo {volume} {595}},\ \bibinfo {pages}
  {378} (\bibinfo {year} {2021})}\BibitemShut {NoStop}%
\bibitem [{\citenamefont {Magrini}\ \emph {et~al.}(2021)\citenamefont
  {Magrini}, \citenamefont {Rosenzweig}, \citenamefont {Bach}, \citenamefont
  {Deutschmann-Olek}, \citenamefont {Hofer}, \citenamefont {Hong},
  \citenamefont {Kiesel}, \citenamefont {Kugi},\ and\ \citenamefont
  {Aspelmeyer}}]{Magrini2021}%
  \BibitemOpen
  \bibfield  {author} {\bibinfo {author} {\bibfnamefont {L.}~\bibnamefont
  {Magrini}}, \bibinfo {author} {\bibfnamefont {P.}~\bibnamefont {Rosenzweig}},
  \bibinfo {author} {\bibfnamefont {C.}~\bibnamefont {Bach}}, \bibinfo {author}
  {\bibfnamefont {A.}~\bibnamefont {Deutschmann-Olek}}, \bibinfo {author}
  {\bibfnamefont {S.~G.}\ \bibnamefont {Hofer}}, \bibinfo {author}
  {\bibfnamefont {S.}~\bibnamefont {Hong}}, \bibinfo {author} {\bibfnamefont
  {N.}~\bibnamefont {Kiesel}}, \bibinfo {author} {\bibfnamefont
  {A.}~\bibnamefont {Kugi}}, \ and\ \bibinfo {author} {\bibfnamefont
  {M.}~\bibnamefont {Aspelmeyer}},\ }\href {\doibase
  10.1038/s41586-021-03602-3} {\bibfield  {journal} {\bibinfo  {journal}
  {Nature}\ }\textbf {\bibinfo {volume} {595}},\ \bibinfo {pages} {373}
  (\bibinfo {year} {2021})}\BibitemShut {NoStop}%
\bibitem [{\citenamefont {Monteiro}\ \emph {et~al.}(2017)\citenamefont
  {Monteiro}, \citenamefont {Ghosh}, \citenamefont {Fine},\ and\ \citenamefont
  {Moore}}]{Monteiro2017}%
  \BibitemOpen
  \bibfield  {author} {\bibinfo {author} {\bibfnamefont {F.}~\bibnamefont
  {Monteiro}}, \bibinfo {author} {\bibfnamefont {S.}~\bibnamefont {Ghosh}},
  \bibinfo {author} {\bibfnamefont {A.~G.}\ \bibnamefont {Fine}}, \ and\
  \bibinfo {author} {\bibfnamefont {D.~C.}\ \bibnamefont {Moore}},\ }\href
  {\doibase 10.1103/PhysRevA.96.063841} {\bibfield  {journal} {\bibinfo
  {journal} {Phys. Rev. A}\ }\textbf {\bibinfo {volume} {96}},\ \bibinfo
  {pages} {063841} (\bibinfo {year} {2017})}\BibitemShut {NoStop}%
\bibitem [{\citenamefont {Timberlake}\ \emph {et~al.}(2019)\citenamefont
  {Timberlake}, \citenamefont {Gasbarri}, \citenamefont {Vinante},
  \citenamefont {Setter},\ and\ \citenamefont {Ulbricht}}]{Timberlake2019}%
  \BibitemOpen
  \bibfield  {author} {\bibinfo {author} {\bibfnamefont {C.}~\bibnamefont
  {Timberlake}}, \bibinfo {author} {\bibfnamefont {G.}~\bibnamefont
  {Gasbarri}}, \bibinfo {author} {\bibfnamefont {A.}~\bibnamefont {Vinante}},
  \bibinfo {author} {\bibfnamefont {A.}~\bibnamefont {Setter}}, \ and\ \bibinfo
  {author} {\bibfnamefont {H.}~\bibnamefont {Ulbricht}},\ }\href {\doibase
  10.1063/1.5129145} {\bibfield  {journal} {\bibinfo  {journal} {Appl. Phys.
  Lett.}\ }\textbf {\bibinfo {volume} {115}},\ \bibinfo {pages} {224101}
  (\bibinfo {year} {2019})}\BibitemShut {NoStop}%
\bibitem [{\citenamefont {Monteiro}\ \emph {et~al.}(2020)\citenamefont
  {Monteiro}, \citenamefont {Li}, \citenamefont {Afek}, \citenamefont {Li},
  \citenamefont {Mossman},\ and\ \citenamefont {Moore}}]{Monteiro2020}%
  \BibitemOpen
  \bibfield  {author} {\bibinfo {author} {\bibfnamefont {F.}~\bibnamefont
  {Monteiro}}, \bibinfo {author} {\bibfnamefont {W.}~\bibnamefont {Li}},
  \bibinfo {author} {\bibfnamefont {G.}~\bibnamefont {Afek}}, \bibinfo {author}
  {\bibfnamefont {C.-l.}\ \bibnamefont {Li}}, \bibinfo {author} {\bibfnamefont
  {M.}~\bibnamefont {Mossman}}, \ and\ \bibinfo {author} {\bibfnamefont
  {D.~C.}\ \bibnamefont {Moore}},\ }\href {\doibase
  10.1103/PhysRevA.101.053835} {\bibfield  {journal} {\bibinfo  {journal}
  {Phys. Rev. A}\ }\textbf {\bibinfo {volume} {101}},\ \bibinfo {pages}
  {053835} (\bibinfo {year} {2020})}\BibitemShut {NoStop}%
\bibitem [{\citenamefont {Ahn}\ \emph {et~al.}(2020)\citenamefont {Ahn},
  \citenamefont {Xu}, \citenamefont {Bang}, \citenamefont {Ju}, \citenamefont
  {Gao},\ and\ \citenamefont {Li}}]{Ahn2020}%
  \BibitemOpen
  \bibfield  {author} {\bibinfo {author} {\bibfnamefont {J.}~\bibnamefont
  {Ahn}}, \bibinfo {author} {\bibfnamefont {Z.}~\bibnamefont {Xu}}, \bibinfo
  {author} {\bibfnamefont {J.}~\bibnamefont {Bang}}, \bibinfo {author}
  {\bibfnamefont {P.}~\bibnamefont {Ju}}, \bibinfo {author} {\bibfnamefont
  {X.}~\bibnamefont {Gao}}, \ and\ \bibinfo {author} {\bibfnamefont
  {T.}~\bibnamefont {Li}},\ }\href {\doibase 10.1038/s41565-019-0605-9}
  {\bibfield  {journal} {\bibinfo  {journal} {Nature Nanotechnology}\ }\textbf
  {\bibinfo {volume} {15}},\ \bibinfo {pages} {89} (\bibinfo {year}
  {2020})}\BibitemShut {NoStop}%
\bibitem [{\citenamefont {van~der Laan}\ \emph {et~al.}(2021)\citenamefont
  {van~der Laan}, \citenamefont {Tebbenjohanns}, \citenamefont {Reimann},
  \citenamefont {Vijayan}, \citenamefont {Novotny},\ and\ \citenamefont
  {Frimmer}}]{Fons2021}%
  \BibitemOpen
  \bibfield  {author} {\bibinfo {author} {\bibfnamefont {F.}~\bibnamefont
  {van~der Laan}}, \bibinfo {author} {\bibfnamefont {F.}~\bibnamefont
  {Tebbenjohanns}}, \bibinfo {author} {\bibfnamefont {R.}~\bibnamefont
  {Reimann}}, \bibinfo {author} {\bibfnamefont {J.}~\bibnamefont {Vijayan}},
  \bibinfo {author} {\bibfnamefont {L.}~\bibnamefont {Novotny}}, \ and\
  \bibinfo {author} {\bibfnamefont {M.}~\bibnamefont {Frimmer}},\ }\href
  {\doibase 10.1103/PhysRevLett.127.123605} {\bibfield  {journal} {\bibinfo
  {journal} {Phys. Rev. Lett.}\ }\textbf {\bibinfo {volume} {127}},\ \bibinfo
  {pages} {123605} (\bibinfo {year} {2021})}\BibitemShut {NoStop}%
\bibitem [{\citenamefont {Ranjit}\ \emph {et~al.}(2016)\citenamefont {Ranjit},
  \citenamefont {Cunningham}, \citenamefont {Casey},\ and\ \citenamefont
  {Geraci}}]{Gambhir2016}%
  \BibitemOpen
  \bibfield  {author} {\bibinfo {author} {\bibfnamefont {G.}~\bibnamefont
  {Ranjit}}, \bibinfo {author} {\bibfnamefont {M.}~\bibnamefont {Cunningham}},
  \bibinfo {author} {\bibfnamefont {K.}~\bibnamefont {Casey}}, \ and\ \bibinfo
  {author} {\bibfnamefont {A.~A.}\ \bibnamefont {Geraci}},\ }\href {\doibase
  10.1103/PhysRevA.93.053801} {\bibfield  {journal} {\bibinfo  {journal} {Phys.
  Rev. A}\ }\textbf {\bibinfo {volume} {93}},\ \bibinfo {pages} {053801}
  (\bibinfo {year} {2016})}\BibitemShut {NoStop}%
\bibitem [{\citenamefont {Hempston}\ \emph {et~al.}(2017)\citenamefont
  {Hempston}, \citenamefont {Vovrosh}, \citenamefont {Toro{\v{s}}},
  \citenamefont {Winstone}, \citenamefont {Rashid},\ and\ \citenamefont
  {Ulbricht}}]{Hempston2017}%
  \BibitemOpen
  \bibfield  {author} {\bibinfo {author} {\bibfnamefont {D.}~\bibnamefont
  {Hempston}}, \bibinfo {author} {\bibfnamefont {J.}~\bibnamefont {Vovrosh}},
  \bibinfo {author} {\bibfnamefont {M.}~\bibnamefont {Toro{\v{s}}}}, \bibinfo
  {author} {\bibfnamefont {G.}~\bibnamefont {Winstone}}, \bibinfo {author}
  {\bibfnamefont {M.}~\bibnamefont {Rashid}}, \ and\ \bibinfo {author}
  {\bibfnamefont {H.}~\bibnamefont {Ulbricht}},\ }\href {\doibase
  10.1063/1.4993555} {\bibfield  {journal} {\bibinfo  {journal} {Appl. Phys.
  Lett.}\ }\textbf {\bibinfo {volume} {111}},\ \bibinfo {pages} {1} (\bibinfo
  {year} {2017})}\BibitemShut {NoStop}%
\bibitem [{\citenamefont {Hebestreit}\ \emph
  {et~al.}(2018{\natexlab{a}})\citenamefont {Hebestreit}, \citenamefont
  {Frimmer}, \citenamefont {Reimann},\ and\ \citenamefont
  {Novotny}}]{Hebestreit2018}%
  \BibitemOpen
  \bibfield  {author} {\bibinfo {author} {\bibfnamefont {E.}~\bibnamefont
  {Hebestreit}}, \bibinfo {author} {\bibfnamefont {M.}~\bibnamefont {Frimmer}},
  \bibinfo {author} {\bibfnamefont {R.}~\bibnamefont {Reimann}}, \ and\
  \bibinfo {author} {\bibfnamefont {L.}~\bibnamefont {Novotny}},\ }\href
  {\doibase 10.1103/PhysRevLett.121.063602} {\bibfield  {journal} {\bibinfo
  {journal} {Phys. Rev. Lett.}\ }\textbf {\bibinfo {volume} {121}},\ \bibinfo
  {pages} {063602} (\bibinfo {year} {2018}{\natexlab{a}})}\BibitemShut
  {NoStop}%
\bibitem [{\citenamefont {Chauhan}\ \emph {et~al.}(2020)\citenamefont
  {Chauhan}, \citenamefont {{\v{C}}ernot{\'i}k},\ and\ \citenamefont
  {Filip}}]{Chauhan2020}%
  \BibitemOpen
  \bibfield  {author} {\bibinfo {author} {\bibfnamefont {A.~K.}\ \bibnamefont
  {Chauhan}}, \bibinfo {author} {\bibfnamefont {O.}~\bibnamefont
  {{\v{C}}ernot{\'i}k}}, \ and\ \bibinfo {author} {\bibfnamefont
  {R.}~\bibnamefont {Filip}},\ }\href {\doibase 10.1088/1367-2630/abcce6}
  {\bibfield  {journal} {\bibinfo  {journal} {New Journal of Physics}\ }\textbf
  {\bibinfo {volume} {22}},\ \bibinfo {pages} {123021} (\bibinfo {year}
  {2020})}\BibitemShut {NoStop}%
\bibitem [{\citenamefont {Brand{\~a}o}\ \emph {et~al.}(2021)\citenamefont
  {Brand{\~a}o}, \citenamefont {Tandeitnik},\ and\ \citenamefont
  {T}}]{Brandao2021}%
  \BibitemOpen
  \bibfield  {author} {\bibinfo {author} {\bibfnamefont {I.}~\bibnamefont
  {Brand{\~a}o}}, \bibinfo {author} {\bibfnamefont {D.}~\bibnamefont
  {Tandeitnik}}, \ and\ \bibinfo {author} {\bibfnamefont {G.}~\bibnamefont
  {T}},\ }\href {\doibase 10.1088/2058-9565/ac1a01} {\bibfield  {journal}
  {\bibinfo  {journal} {Quantum Science and Technology}\ }\textbf {\bibinfo
  {volume} {6}},\ \bibinfo {pages} {045013} (\bibinfo {year}
  {2021})}\BibitemShut {NoStop}%
\bibitem [{\citenamefont {Kotler}\ \emph {et~al.}(2021)\citenamefont {Kotler},
  \citenamefont {Peterson}, \citenamefont {Shojaee}, \citenamefont {Lecocq},
  \citenamefont {Cicak}, \citenamefont {Kwiatkowski}, \citenamefont {Geller},
  \citenamefont {Glancy}, \citenamefont {Knill}, \citenamefont {Simmonds},
  \citenamefont {Aumentado},\ and\ \citenamefont {Teufel}}]{Kotler2021}%
  \BibitemOpen
  \bibfield  {author} {\bibinfo {author} {\bibfnamefont {S.}~\bibnamefont
  {Kotler}}, \bibinfo {author} {\bibfnamefont {G.~A.}\ \bibnamefont
  {Peterson}}, \bibinfo {author} {\bibfnamefont {E.}~\bibnamefont {Shojaee}},
  \bibinfo {author} {\bibfnamefont {F.}~\bibnamefont {Lecocq}}, \bibinfo
  {author} {\bibfnamefont {K.}~\bibnamefont {Cicak}}, \bibinfo {author}
  {\bibfnamefont {A.}~\bibnamefont {Kwiatkowski}}, \bibinfo {author}
  {\bibfnamefont {S.}~\bibnamefont {Geller}}, \bibinfo {author} {\bibfnamefont
  {S.}~\bibnamefont {Glancy}}, \bibinfo {author} {\bibfnamefont
  {E.}~\bibnamefont {Knill}}, \bibinfo {author} {\bibfnamefont {R.~W.}\
  \bibnamefont {Simmonds}}, \bibinfo {author} {\bibfnamefont {J.}~\bibnamefont
  {Aumentado}}, \ and\ \bibinfo {author} {\bibfnamefont {J.~D.}\ \bibnamefont
  {Teufel}},\ }\href {\doibase 10.1126/science.abf2998} {\bibfield  {journal}
  {\bibinfo  {journal} {Science}\ }\textbf {\bibinfo {volume} {372}},\ \bibinfo
  {pages} {622} (\bibinfo {year} {2021})}\BibitemShut {NoStop}%
\bibitem [{\citenamefont {de~Lépinay}\ \emph {et~al.}(2021)\citenamefont
  {de~Lépinay}, \citenamefont {Ockeloen-Korppi}, \citenamefont {Woolley},\
  and\ \citenamefont {Sillanpää}}]{Lepinay2021}%
  \BibitemOpen
  \bibfield  {author} {\bibinfo {author} {\bibfnamefont {L.~M.}\ \bibnamefont
  {de~Lépinay}}, \bibinfo {author} {\bibfnamefont {C.~F.}\ \bibnamefont
  {Ockeloen-Korppi}}, \bibinfo {author} {\bibfnamefont {M.~J.}\ \bibnamefont
  {Woolley}}, \ and\ \bibinfo {author} {\bibfnamefont {M.~A.}\ \bibnamefont
  {Sillanpää}},\ }\href {\doibase 10.1126/science.abf5389} {\bibfield
  {journal} {\bibinfo  {journal} {Science}\ }\textbf {\bibinfo {volume}
  {372}},\ \bibinfo {pages} {625} (\bibinfo {year} {2021})}\BibitemShut
  {NoStop}%
\bibitem [{\citenamefont {Reimann}\ \emph {et~al.}(2015)\citenamefont
  {Reimann}, \citenamefont {Alt}, \citenamefont {Kampschulte}, \citenamefont
  {Macha}, \citenamefont {Ratschbacher}, \citenamefont {Thau}, \citenamefont
  {Yoon},\ and\ \citenamefont {Meschede}}]{Reimann2015}%
  \BibitemOpen
  \bibfield  {author} {\bibinfo {author} {\bibfnamefont {R.}~\bibnamefont
  {Reimann}}, \bibinfo {author} {\bibfnamefont {W.}~\bibnamefont {Alt}},
  \bibinfo {author} {\bibfnamefont {T.}~\bibnamefont {Kampschulte}}, \bibinfo
  {author} {\bibfnamefont {T.}~\bibnamefont {Macha}}, \bibinfo {author}
  {\bibfnamefont {L.}~\bibnamefont {Ratschbacher}}, \bibinfo {author}
  {\bibfnamefont {N.}~\bibnamefont {Thau}}, \bibinfo {author} {\bibfnamefont
  {S.}~\bibnamefont {Yoon}}, \ and\ \bibinfo {author} {\bibfnamefont
  {D.}~\bibnamefont {Meschede}},\ }\href {\doibase
  10.1103/PhysRevLett.114.023601} {\bibfield  {journal} {\bibinfo  {journal}
  {Phys. Rev. Lett.}\ }\textbf {\bibinfo {volume} {114}},\ \bibinfo {pages}
  {023601} (\bibinfo {year} {2015})}\BibitemShut {NoStop}%
\bibitem [{\citenamefont {Landig}\ \emph {et~al.}(2016)\citenamefont {Landig},
  \citenamefont {Hruby}, \citenamefont {Dogra}, \citenamefont {Landini},
  \citenamefont {Mottl}, \citenamefont {Donner},\ and\ \citenamefont
  {Esslinger}}]{Landig2016}%
  \BibitemOpen
  \bibfield  {author} {\bibinfo {author} {\bibfnamefont {R.}~\bibnamefont
  {Landig}}, \bibinfo {author} {\bibfnamefont {L.}~\bibnamefont {Hruby}},
  \bibinfo {author} {\bibfnamefont {N.}~\bibnamefont {Dogra}}, \bibinfo
  {author} {\bibfnamefont {M.}~\bibnamefont {Landini}}, \bibinfo {author}
  {\bibfnamefont {R.}~\bibnamefont {Mottl}}, \bibinfo {author} {\bibfnamefont
  {T.}~\bibnamefont {Donner}}, \ and\ \bibinfo {author} {\bibfnamefont
  {T.}~\bibnamefont {Esslinger}},\ }\href {\doibase 10.1038/nature17409}
  {\bibfield  {journal} {\bibinfo  {journal} {Nature}\ }\textbf {\bibinfo
  {volume} {532}},\ \bibinfo {pages} {476} (\bibinfo {year}
  {2016})}\BibitemShut {NoStop}%
\bibitem [{\citenamefont {Bernien}\ \emph {et~al.}(2017)\citenamefont
  {Bernien}, \citenamefont {Schwartz}, \citenamefont {Keesling}, \citenamefont
  {Levine}, \citenamefont {Omran}, \citenamefont {Pichler}, \citenamefont
  {Choi}, \citenamefont {Zibrov}, \citenamefont {Endres}, \citenamefont
  {Greiner}, \citenamefont {Vuleti{\'{c}}},\ and\ \citenamefont
  {Lukin}}]{Bernien2017}%
  \BibitemOpen
  \bibfield  {author} {\bibinfo {author} {\bibfnamefont {H.}~\bibnamefont
  {Bernien}}, \bibinfo {author} {\bibfnamefont {S.}~\bibnamefont {Schwartz}},
  \bibinfo {author} {\bibfnamefont {A.}~\bibnamefont {Keesling}}, \bibinfo
  {author} {\bibfnamefont {H.}~\bibnamefont {Levine}}, \bibinfo {author}
  {\bibfnamefont {A.}~\bibnamefont {Omran}}, \bibinfo {author} {\bibfnamefont
  {H.}~\bibnamefont {Pichler}}, \bibinfo {author} {\bibfnamefont
  {S.}~\bibnamefont {Choi}}, \bibinfo {author} {\bibfnamefont {A.~S.}\
  \bibnamefont {Zibrov}}, \bibinfo {author} {\bibfnamefont {M.}~\bibnamefont
  {Endres}}, \bibinfo {author} {\bibfnamefont {M.}~\bibnamefont {Greiner}},
  \bibinfo {author} {\bibfnamefont {V.}~\bibnamefont {Vuleti{\'{c}}}}, \ and\
  \bibinfo {author} {\bibfnamefont {M.~D.}\ \bibnamefont {Lukin}},\ }\href
  {\doibase 10.1038/nature24622} {\bibfield  {journal} {\bibinfo  {journal}
  {Nature}\ }\textbf {\bibinfo {volume} {551}},\ \bibinfo {pages} {579}
  (\bibinfo {year} {2017})}\BibitemShut {NoStop}%
\bibitem [{\citenamefont {Liu}\ \emph {et~al.}(2020)\citenamefont {Liu},
  \citenamefont {Yin},\ and\ \citenamefont {Li}}]{Liu2020}%
  \BibitemOpen
  \bibfield  {author} {\bibinfo {author} {\bibfnamefont {S.}~\bibnamefont
  {Liu}}, \bibinfo {author} {\bibfnamefont {Z.-q.}\ \bibnamefont {Yin}}, \ and\
  \bibinfo {author} {\bibfnamefont {T.}~\bibnamefont {Li}},\ }\href {\doibase
  10.1002/qute.201900099} {\bibfield  {journal} {\bibinfo  {journal} {Advanced
  Quantum Technologies}\ }\textbf {\bibinfo {volume} {3}},\ \bibinfo {pages}
  {1900099} (\bibinfo {year} {2020})}\BibitemShut {NoStop}%
\bibitem [{\citenamefont {Periwal}\ \emph {et~al.}(2021)\citenamefont
  {Periwal}, \citenamefont {Cooper}, \citenamefont {Kunkel}, \citenamefont
  {Wienand}, \citenamefont {Davis},\ and\ \citenamefont
  {Schleier-Smith}}]{Periwal2021}%
  \BibitemOpen
  \bibfield  {author} {\bibinfo {author} {\bibfnamefont {A.}~\bibnamefont
  {Periwal}}, \bibinfo {author} {\bibfnamefont {E.~S.}\ \bibnamefont {Cooper}},
  \bibinfo {author} {\bibfnamefont {P.}~\bibnamefont {Kunkel}}, \bibinfo
  {author} {\bibfnamefont {J.~F.}\ \bibnamefont {Wienand}}, \bibinfo {author}
  {\bibfnamefont {E.~J.}\ \bibnamefont {Davis}}, \ and\ \bibinfo {author}
  {\bibfnamefont {M.}~\bibnamefont {Schleier-Smith}},\ }\href {\doibase
  10.1038/s41586-021-04156-0} {\bibfield  {journal} {\bibinfo  {journal}
  {Nature}\ }\textbf {\bibinfo {volume} {600}},\ \bibinfo {pages} {630}
  (\bibinfo {year} {2021})}\BibitemShut {NoStop}%
\bibitem [{\citenamefont {Bressi}\ \emph {et~al.}(2002)\citenamefont {Bressi},
  \citenamefont {Carugno}, \citenamefont {Onofrio},\ and\ \citenamefont
  {Ruoso}}]{Bressi2002}%
  \BibitemOpen
  \bibfield  {author} {\bibinfo {author} {\bibfnamefont {G.}~\bibnamefont
  {Bressi}}, \bibinfo {author} {\bibfnamefont {G.}~\bibnamefont {Carugno}},
  \bibinfo {author} {\bibfnamefont {R.}~\bibnamefont {Onofrio}}, \ and\
  \bibinfo {author} {\bibfnamefont {G.}~\bibnamefont {Ruoso}},\ }\href
  {\doibase 10.1103/PhysRevLett.88.041804} {\bibfield  {journal} {\bibinfo
  {journal} {Phys. Rev. Lett.}\ }\textbf {\bibinfo {volume} {88}},\ \bibinfo
  {pages} {041804} (\bibinfo {year} {2002})}\BibitemShut {NoStop}%
\bibitem [{\citenamefont {Wang}\ \emph {et~al.}(2021)\citenamefont {Wang},
  \citenamefont {Tang}, \citenamefont {Ng}, \citenamefont {Messina},
  \citenamefont {Guizal}, \citenamefont {Crosse}, \citenamefont {Antezza},
  \citenamefont {Chan},\ and\ \citenamefont {Chan}}]{Wang2021}%
  \BibitemOpen
  \bibfield  {author} {\bibinfo {author} {\bibfnamefont {M.}~\bibnamefont
  {Wang}}, \bibinfo {author} {\bibfnamefont {L.}~\bibnamefont {Tang}}, \bibinfo
  {author} {\bibfnamefont {C.~Y.}\ \bibnamefont {Ng}}, \bibinfo {author}
  {\bibfnamefont {R.}~\bibnamefont {Messina}}, \bibinfo {author} {\bibfnamefont
  {B.}~\bibnamefont {Guizal}}, \bibinfo {author} {\bibfnamefont {J.~A.}\
  \bibnamefont {Crosse}}, \bibinfo {author} {\bibfnamefont {M.}~\bibnamefont
  {Antezza}}, \bibinfo {author} {\bibfnamefont {C.~T.}\ \bibnamefont {Chan}}, \
  and\ \bibinfo {author} {\bibfnamefont {H.~B.}\ \bibnamefont {Chan}},\ }\href
  {\doibase 10.1038/s41467-021-20891-4} {\bibfield  {journal} {\bibinfo
  {journal} {Nature Communications}\ }\textbf {\bibinfo {volume} {12}},\
  \bibinfo {pages} {600} (\bibinfo {year} {2021})}\BibitemShut {NoStop}%
\bibitem [{\citenamefont {Quinn}\ \emph {et~al.}(2001)\citenamefont {Quinn},
  \citenamefont {Speake}, \citenamefont {Richman}, \citenamefont {Davis},\ and\
  \citenamefont {Picard}}]{Quinn2001}%
  \BibitemOpen
  \bibfield  {author} {\bibinfo {author} {\bibfnamefont {T.~J.}\ \bibnamefont
  {Quinn}}, \bibinfo {author} {\bibfnamefont {C.~C.}\ \bibnamefont {Speake}},
  \bibinfo {author} {\bibfnamefont {S.~J.}\ \bibnamefont {Richman}}, \bibinfo
  {author} {\bibfnamefont {R.~S.}\ \bibnamefont {Davis}}, \ and\ \bibinfo
  {author} {\bibfnamefont {A.}~\bibnamefont {Picard}},\ }\href {\doibase
  10.1103/PhysRevLett.87.111101} {\bibfield  {journal} {\bibinfo  {journal}
  {Phys. Rev. Lett.}\ }\textbf {\bibinfo {volume} {87}},\ \bibinfo {pages}
  {111101} (\bibinfo {year} {2001})}\BibitemShut {NoStop}%
\bibitem [{\citenamefont {Li}\ \emph {et~al.}(2018)\citenamefont {Li},
  \citenamefont {Xue}, \citenamefont {Liu}, \citenamefont {Wu}, \citenamefont
  {Yang}, \citenamefont {Shao}, \citenamefont {Quan}, \citenamefont {Tan},
  \citenamefont {Tu}, \citenamefont {Liu}, \citenamefont {Xu}, \citenamefont
  {Liu}, \citenamefont {Wang}, \citenamefont {Hu}, \citenamefont {Zhou},
  \citenamefont {Luo}, \citenamefont {Wu}, \citenamefont {Milyukov},\ and\
  \citenamefont {Luo}}]{Li2018}%
  \BibitemOpen
  \bibfield  {author} {\bibinfo {author} {\bibfnamefont {Q.}~\bibnamefont
  {Li}}, \bibinfo {author} {\bibfnamefont {C.}~\bibnamefont {Xue}}, \bibinfo
  {author} {\bibfnamefont {J.-P.}\ \bibnamefont {Liu}}, \bibinfo {author}
  {\bibfnamefont {J.-F.}\ \bibnamefont {Wu}}, \bibinfo {author} {\bibfnamefont
  {S.-Q.}\ \bibnamefont {Yang}}, \bibinfo {author} {\bibfnamefont {C.-G.}\
  \bibnamefont {Shao}}, \bibinfo {author} {\bibfnamefont {L.-D.}\ \bibnamefont
  {Quan}}, \bibinfo {author} {\bibfnamefont {W.-H.}\ \bibnamefont {Tan}},
  \bibinfo {author} {\bibfnamefont {L.-C.}\ \bibnamefont {Tu}}, \bibinfo
  {author} {\bibfnamefont {Q.}~\bibnamefont {Liu}}, \bibinfo {author}
  {\bibfnamefont {H.}~\bibnamefont {Xu}}, \bibinfo {author} {\bibfnamefont
  {L.-X.}\ \bibnamefont {Liu}}, \bibinfo {author} {\bibfnamefont {Q.-L.}\
  \bibnamefont {Wang}}, \bibinfo {author} {\bibfnamefont {Z.-K.}\ \bibnamefont
  {Hu}}, \bibinfo {author} {\bibfnamefont {Z.-B.}\ \bibnamefont {Zhou}},
  \bibinfo {author} {\bibfnamefont {P.-S.}\ \bibnamefont {Luo}}, \bibinfo
  {author} {\bibfnamefont {S.-C.}\ \bibnamefont {Wu}}, \bibinfo {author}
  {\bibfnamefont {V.}~\bibnamefont {Milyukov}}, \ and\ \bibinfo {author}
  {\bibfnamefont {J.}~\bibnamefont {Luo}},\ }\href {\doibase
  10.1038/s41586-018-0431-5} {\bibfield  {journal} {\bibinfo  {journal}
  {Nature}\ }\textbf {\bibinfo {volume} {560}},\ \bibinfo {pages} {582}
  (\bibinfo {year} {2018})}\BibitemShut {NoStop}%
\bibitem [{\citenamefont {Kaufman}\ \emph {et~al.}(2014)\citenamefont
  {Kaufman}, \citenamefont {Lester}, \citenamefont {Reynolds}, \citenamefont
  {Wall}, \citenamefont {Foss-Feig}, \citenamefont {Hazzard}, \citenamefont
  {Rey},\ and\ \citenamefont {Regal}}]{Kaufman2014}%
  \BibitemOpen
  \bibfield  {author} {\bibinfo {author} {\bibfnamefont {A.~M.}\ \bibnamefont
  {Kaufman}}, \bibinfo {author} {\bibfnamefont {B.~J.}\ \bibnamefont {Lester}},
  \bibinfo {author} {\bibfnamefont {C.~M.}\ \bibnamefont {Reynolds}}, \bibinfo
  {author} {\bibfnamefont {M.~L.}\ \bibnamefont {Wall}}, \bibinfo {author}
  {\bibfnamefont {M.}~\bibnamefont {Foss-Feig}}, \bibinfo {author}
  {\bibfnamefont {K.~R.~A.}\ \bibnamefont {Hazzard}}, \bibinfo {author}
  {\bibfnamefont {A.~M.}\ \bibnamefont {Rey}}, \ and\ \bibinfo {author}
  {\bibfnamefont {C.~A.}\ \bibnamefont {Regal}},\ }\href {\doibase
  10.1126/science.1250057} {\bibfield  {journal} {\bibinfo  {journal}
  {Science}\ }\textbf {\bibinfo {volume} {345}},\ \bibinfo {pages} {306}
  (\bibinfo {year} {2014})}\BibitemShut {NoStop}%
\bibitem [{\citenamefont {Endres}\ \emph {et~al.}(2016)\citenamefont {Endres},
  \citenamefont {Bernien}, \citenamefont {Keesling}, \citenamefont {Levine},
  \citenamefont {Anschuetz}, \citenamefont {Krajenbrink}, \citenamefont
  {Senko}, \citenamefont {Vuletic}, \citenamefont {Greiner},\ and\
  \citenamefont {Lukin}}]{Endres2016}%
  \BibitemOpen
  \bibfield  {author} {\bibinfo {author} {\bibfnamefont {M.}~\bibnamefont
  {Endres}}, \bibinfo {author} {\bibfnamefont {H.}~\bibnamefont {Bernien}},
  \bibinfo {author} {\bibfnamefont {A.}~\bibnamefont {Keesling}}, \bibinfo
  {author} {\bibfnamefont {H.}~\bibnamefont {Levine}}, \bibinfo {author}
  {\bibfnamefont {E.~R.}\ \bibnamefont {Anschuetz}}, \bibinfo {author}
  {\bibfnamefont {A.}~\bibnamefont {Krajenbrink}}, \bibinfo {author}
  {\bibfnamefont {C.}~\bibnamefont {Senko}}, \bibinfo {author} {\bibfnamefont
  {V.}~\bibnamefont {Vuletic}}, \bibinfo {author} {\bibfnamefont
  {M.}~\bibnamefont {Greiner}}, \ and\ \bibinfo {author} {\bibfnamefont
  {M.~D.}\ \bibnamefont {Lukin}},\ }\href {\doibase 10.1126/science.aah3752}
  {\bibfield  {journal} {\bibinfo  {journal} {Science}\ }\textbf {\bibinfo
  {volume} {354}},\ \bibinfo {pages} {1024} (\bibinfo {year}
  {2016})}\BibitemShut {NoStop}%
\bibitem [{\citenamefont {Barredo}\ \emph {et~al.}(2016)\citenamefont
  {Barredo}, \citenamefont {de~L{\'e}s{\'e}leuc}, \citenamefont {Lienhard},
  \citenamefont {Lahaye},\ and\ \citenamefont {Browaeys}}]{Daniel2016}%
  \BibitemOpen
  \bibfield  {author} {\bibinfo {author} {\bibfnamefont {D.}~\bibnamefont
  {Barredo}}, \bibinfo {author} {\bibfnamefont {S.}~\bibnamefont
  {de~L{\'e}s{\'e}leuc}}, \bibinfo {author} {\bibfnamefont {V.}~\bibnamefont
  {Lienhard}}, \bibinfo {author} {\bibfnamefont {T.}~\bibnamefont {Lahaye}}, \
  and\ \bibinfo {author} {\bibfnamefont {A.}~\bibnamefont {Browaeys}},\ }\href
  {\doibase 10.1126/science.aah3778} {\bibfield  {journal} {\bibinfo  {journal}
  {Science}\ }\textbf {\bibinfo {volume} {354}},\ \bibinfo {pages} {1021}
  (\bibinfo {year} {2016})}\BibitemShut {NoStop}%
\bibitem [{\citenamefont {Ebadi}\ \emph {et~al.}(2021)\citenamefont {Ebadi},
  \citenamefont {Wang}, \citenamefont {Levine}, \citenamefont {Keesling},
  \citenamefont {Semeghini}, \citenamefont {Omran}, \citenamefont {Bluvstein},
  \citenamefont {Samajdar}, \citenamefont {Pichler}, \citenamefont {Ho},
  \citenamefont {Choi}, \citenamefont {Sachdev}, \citenamefont {Greiner},
  \citenamefont {Vuletic},\ and\ \citenamefont {Lukin}}]{Ebadi2021}%
  \BibitemOpen
  \bibfield  {author} {\bibinfo {author} {\bibfnamefont {S.}~\bibnamefont
  {Ebadi}}, \bibinfo {author} {\bibfnamefont {T.~T.}\ \bibnamefont {Wang}},
  \bibinfo {author} {\bibfnamefont {H.}~\bibnamefont {Levine}}, \bibinfo
  {author} {\bibfnamefont {A.}~\bibnamefont {Keesling}}, \bibinfo {author}
  {\bibfnamefont {G.}~\bibnamefont {Semeghini}}, \bibinfo {author}
  {\bibfnamefont {A.}~\bibnamefont {Omran}}, \bibinfo {author} {\bibfnamefont
  {D.}~\bibnamefont {Bluvstein}}, \bibinfo {author} {\bibfnamefont
  {R.}~\bibnamefont {Samajdar}}, \bibinfo {author} {\bibfnamefont
  {H.}~\bibnamefont {Pichler}}, \bibinfo {author} {\bibfnamefont {W.~W.}\
  \bibnamefont {Ho}}, \bibinfo {author} {\bibfnamefont {S.}~\bibnamefont
  {Choi}}, \bibinfo {author} {\bibfnamefont {S.}~\bibnamefont {Sachdev}},
  \bibinfo {author} {\bibfnamefont {M.}~\bibnamefont {Greiner}}, \bibinfo
  {author} {\bibfnamefont {V.}~\bibnamefont {Vuletic}}, \ and\ \bibinfo
  {author} {\bibfnamefont {M.~D.}\ \bibnamefont {Lukin}},\ }\href {\doibase
  10.1038/s41586-021-03582-4} {\bibfield  {journal} {\bibinfo  {journal}
  {Nature}\ }\textbf {\bibinfo {volume} {595}},\ \bibinfo {pages} {227}
  (\bibinfo {year} {2021})}\BibitemShut {NoStop}%
\bibitem [{\citenamefont {Dania}\ \emph {et~al.}(2021)\citenamefont {Dania},
  \citenamefont {Bykov}, \citenamefont {Knoll}, \citenamefont {Mestres},\ and\
  \citenamefont {Northup}}]{Dania2021}%
  \BibitemOpen
  \bibfield  {author} {\bibinfo {author} {\bibfnamefont {L.}~\bibnamefont
  {Dania}}, \bibinfo {author} {\bibfnamefont {D.~S.}\ \bibnamefont {Bykov}},
  \bibinfo {author} {\bibfnamefont {M.}~\bibnamefont {Knoll}}, \bibinfo
  {author} {\bibfnamefont {P.}~\bibnamefont {Mestres}}, \ and\ \bibinfo
  {author} {\bibfnamefont {T.~E.}\ \bibnamefont {Northup}},\ }\href {\doibase
  10.1103/PhysRevResearch.3.013018} {\bibfield  {journal} {\bibinfo  {journal}
  {Phys. Rev. Research}\ }\textbf {\bibinfo {volume} {3}},\ \bibinfo {pages}
  {013018} (\bibinfo {year} {2021})}\BibitemShut {NoStop}%
\bibitem [{\citenamefont {Bang}\ \emph {et~al.}(2020)\citenamefont {Bang},
  \citenamefont {Seberson}, \citenamefont {Ju}, \citenamefont {Ahn},
  \citenamefont {Xu}, \citenamefont {Gao}, \citenamefont {Robicheaux},\ and\
  \citenamefont {Li}}]{Bang2020}%
  \BibitemOpen
  \bibfield  {author} {\bibinfo {author} {\bibfnamefont {J.}~\bibnamefont
  {Bang}}, \bibinfo {author} {\bibfnamefont {T.}~\bibnamefont {Seberson}},
  \bibinfo {author} {\bibfnamefont {P.}~\bibnamefont {Ju}}, \bibinfo {author}
  {\bibfnamefont {J.}~\bibnamefont {Ahn}}, \bibinfo {author} {\bibfnamefont
  {Z.}~\bibnamefont {Xu}}, \bibinfo {author} {\bibfnamefont {X.}~\bibnamefont
  {Gao}}, \bibinfo {author} {\bibfnamefont {F.}~\bibnamefont {Robicheaux}}, \
  and\ \bibinfo {author} {\bibfnamefont {T.}~\bibnamefont {Li}},\ }\href
  {\doibase 10.1103/PhysRevResearch.2.043054} {\bibfield  {journal} {\bibinfo
  {journal} {Phys. Rev. Research}\ }\textbf {\bibinfo {volume} {2}},\ \bibinfo
  {pages} {043054} (\bibinfo {year} {2020})}\BibitemShut {NoStop}%
\bibitem [{\citenamefont {Deli{\'{c}}}\ \emph {et~al.}(2019)\citenamefont
  {Deli{\'{c}}}, \citenamefont {Reisenbauer}, \citenamefont {Grass},
  \citenamefont {Kiesel}, \citenamefont {Vuleti{\'{c}}},\ and\ \citenamefont
  {Aspelmeyer}}]{Delic2019}%
  \BibitemOpen
  \bibfield  {author} {\bibinfo {author} {\bibfnamefont {U.}~\bibnamefont
  {Deli{\'{c}}}}, \bibinfo {author} {\bibfnamefont {M.}~\bibnamefont
  {Reisenbauer}}, \bibinfo {author} {\bibfnamefont {D.}~\bibnamefont {Grass}},
  \bibinfo {author} {\bibfnamefont {N.}~\bibnamefont {Kiesel}}, \bibinfo
  {author} {\bibfnamefont {V.}~\bibnamefont {Vuleti{\'{c}}}}, \ and\ \bibinfo
  {author} {\bibfnamefont {M.}~\bibnamefont {Aspelmeyer}},\ }\href {\doibase
  10.1103/PhysRevLett.122.123602} {\bibfield  {journal} {\bibinfo  {journal}
  {Phys. Rev. Lett.}\ }\textbf {\bibinfo {volume} {122}},\ \bibinfo {pages}
  {123602} (\bibinfo {year} {2019})}\BibitemShut {NoStop}%
\bibitem [{\citenamefont {Windey}\ \emph {et~al.}(2019)\citenamefont {Windey},
  \citenamefont {Gonzalez-Ballestero}, \citenamefont {Maurer}, \citenamefont
  {Novotny}, \citenamefont {Romero-Isart},\ and\ \citenamefont
  {Reimann}}]{Windey2019}%
  \BibitemOpen
  \bibfield  {author} {\bibinfo {author} {\bibfnamefont {D.}~\bibnamefont
  {Windey}}, \bibinfo {author} {\bibfnamefont {C.}~\bibnamefont
  {Gonzalez-Ballestero}}, \bibinfo {author} {\bibfnamefont {P.}~\bibnamefont
  {Maurer}}, \bibinfo {author} {\bibfnamefont {L.}~\bibnamefont {Novotny}},
  \bibinfo {author} {\bibfnamefont {O.}~\bibnamefont {Romero-Isart}}, \ and\
  \bibinfo {author} {\bibfnamefont {R.}~\bibnamefont {Reimann}},\ }\href
  {\doibase 10.1103/PhysRevLett.122.123601} {\bibfield  {journal} {\bibinfo
  {journal} {Phys. Rev. Lett.}\ }\textbf {\bibinfo {volume} {122}},\ \bibinfo
  {pages} {123601} (\bibinfo {year} {2019})}\BibitemShut {NoStop}%
\bibitem [{\citenamefont {Meyer}\ \emph {et~al.}(2019)\citenamefont {Meyer},
  \citenamefont {Sommer}, \citenamefont {Mestres}, \citenamefont {Gieseler},
  \citenamefont {Jain}, \citenamefont {Novotny},\ and\ \citenamefont
  {Quidant}}]{Meyer2019}%
  \BibitemOpen
  \bibfield  {author} {\bibinfo {author} {\bibfnamefont {N.}~\bibnamefont
  {Meyer}}, \bibinfo {author} {\bibfnamefont {A.~d. l.~R.}\ \bibnamefont
  {Sommer}}, \bibinfo {author} {\bibfnamefont {P.}~\bibnamefont {Mestres}},
  \bibinfo {author} {\bibfnamefont {J.}~\bibnamefont {Gieseler}}, \bibinfo
  {author} {\bibfnamefont {V.}~\bibnamefont {Jain}}, \bibinfo {author}
  {\bibfnamefont {L.}~\bibnamefont {Novotny}}, \ and\ \bibinfo {author}
  {\bibfnamefont {R.}~\bibnamefont {Quidant}},\ }\href {\doibase
  10.1103/PhysRevLett.123.153601} {\bibfield  {journal} {\bibinfo  {journal}
  {Phys. Rev. Lett.}\ }\textbf {\bibinfo {volume} {123}},\ \bibinfo {pages}
  {153601} (\bibinfo {year} {2019})}\BibitemShut {NoStop}%
\bibitem [{\citenamefont {Wilson}\ \emph {et~al.}(2015)\citenamefont {Wilson},
  \citenamefont {Sudhir}, \citenamefont {Piro}, \citenamefont {Schilling},
  \citenamefont {Ghadimi},\ and\ \citenamefont {Kippenberg}}]{Wilson2015}%
  \BibitemOpen
  \bibfield  {author} {\bibinfo {author} {\bibfnamefont {D.~J.}\ \bibnamefont
  {Wilson}}, \bibinfo {author} {\bibfnamefont {V.}~\bibnamefont {Sudhir}},
  \bibinfo {author} {\bibfnamefont {N.}~\bibnamefont {Piro}}, \bibinfo {author}
  {\bibfnamefont {R.}~\bibnamefont {Schilling}}, \bibinfo {author}
  {\bibfnamefont {A.}~\bibnamefont {Ghadimi}}, \ and\ \bibinfo {author}
  {\bibfnamefont {T.~J.}\ \bibnamefont {Kippenberg}},\ }\href {\doibase
  10.1038/nature14672} {\bibfield  {journal} {\bibinfo  {journal} {Nature}\
  }\textbf {\bibinfo {volume} {524}},\ \bibinfo {pages} {325} (\bibinfo {year}
  {2015})}\BibitemShut {NoStop}%
\bibitem [{\citenamefont {Tebbenjohanns}\ \emph
  {et~al.}(2019{\natexlab{a}})\citenamefont {Tebbenjohanns}, \citenamefont
  {Frimmer}, \citenamefont {Militaru}, \citenamefont {Jain},\ and\
  \citenamefont {Novotny}}]{Tebbenjohanns2019colddamping}%
  \BibitemOpen
  \bibfield  {author} {\bibinfo {author} {\bibfnamefont {F.}~\bibnamefont
  {Tebbenjohanns}}, \bibinfo {author} {\bibfnamefont {M.}~\bibnamefont
  {Frimmer}}, \bibinfo {author} {\bibfnamefont {A.}~\bibnamefont {Militaru}},
  \bibinfo {author} {\bibfnamefont {V.}~\bibnamefont {Jain}}, \ and\ \bibinfo
  {author} {\bibfnamefont {L.}~\bibnamefont {Novotny}},\ }\href {\doibase
  10.1103/PhysRevLett.122.223601} {\bibfield  {journal} {\bibinfo  {journal}
  {Phys. Rev. Lett.}\ }\textbf {\bibinfo {volume} {122}},\ \bibinfo {pages}
  {223601} (\bibinfo {year} {2019}{\natexlab{a}})},\ \Eprint
  {http://arxiv.org/abs/1812.09875} {1812.09875} \BibitemShut {NoStop}%
\bibitem [{\citenamefont {Rudolph}\ \emph {et~al.}(2020)\citenamefont
  {Rudolph}, \citenamefont {Hornberger},\ and\ \citenamefont
  {Stickler}}]{Rudolph2020}%
  \BibitemOpen
  \bibfield  {author} {\bibinfo {author} {\bibfnamefont {H.}~\bibnamefont
  {Rudolph}}, \bibinfo {author} {\bibfnamefont {K.}~\bibnamefont {Hornberger}},
  \ and\ \bibinfo {author} {\bibfnamefont {B.~A.}\ \bibnamefont {Stickler}},\
  }\href {\doibase 10.1103/PhysRevA.101.011804} {\bibfield  {journal} {\bibinfo
   {journal} {Phys. Rev. A}\ }\textbf {\bibinfo {volume} {101}},\ \bibinfo
  {pages} {011804} (\bibinfo {year} {2020})}\BibitemShut {NoStop}%
\bibitem [{\citenamefont {Gieseler}\ \emph {et~al.}(2012)\citenamefont
  {Gieseler}, \citenamefont {Deutsch}, \citenamefont {Quidant},\ and\
  \citenamefont {Novotny}}]{Gieseler2012}%
  \BibitemOpen
  \bibfield  {author} {\bibinfo {author} {\bibfnamefont {J.}~\bibnamefont
  {Gieseler}}, \bibinfo {author} {\bibfnamefont {B.}~\bibnamefont {Deutsch}},
  \bibinfo {author} {\bibfnamefont {R.}~\bibnamefont {Quidant}}, \ and\
  \bibinfo {author} {\bibfnamefont {L.}~\bibnamefont {Novotny}},\ }\href
  {\doibase 10.1103/PhysRevLett.109.103603} {\bibfield  {journal} {\bibinfo
  {journal} {Phys. Rev. Lett.}\ }\textbf {\bibinfo {volume} {109}},\ \bibinfo
  {pages} {103603} (\bibinfo {year} {2012})}\BibitemShut {NoStop}%
\bibitem [{\citenamefont {Jain}\ \emph {et~al.}(2016)\citenamefont {Jain},
  \citenamefont {Gieseler}, \citenamefont {Moritz}, \citenamefont {Dellago},
  \citenamefont {Quidant},\ and\ \citenamefont {Novotny}}]{Jain2016}%
  \BibitemOpen
  \bibfield  {author} {\bibinfo {author} {\bibfnamefont {V.}~\bibnamefont
  {Jain}}, \bibinfo {author} {\bibfnamefont {J.}~\bibnamefont {Gieseler}},
  \bibinfo {author} {\bibfnamefont {C.}~\bibnamefont {Moritz}}, \bibinfo
  {author} {\bibfnamefont {C.}~\bibnamefont {Dellago}}, \bibinfo {author}
  {\bibfnamefont {R.}~\bibnamefont {Quidant}}, \ and\ \bibinfo {author}
  {\bibfnamefont {L.}~\bibnamefont {Novotny}},\ }\href {\doibase
  10.1103/PhysRevLett.116.243601} {\bibfield  {journal} {\bibinfo  {journal}
  {Phys. Rev. Lett.}\ }\textbf {\bibinfo {volume} {116}},\ \bibinfo {pages}
  {243601} (\bibinfo {year} {2016})}\BibitemShut {NoStop}%
\bibitem [{\citenamefont {Rieser}\ \emph {et~al.}(2022)\citenamefont {Rieser},
  \citenamefont {Ciampini}, \citenamefont {Rudolph}, \citenamefont {Kiesel},
  \citenamefont {Hornberger}, \citenamefont {Stickler}, \citenamefont
  {Aspelmeyer},\ and\ \citenamefont {Delić}}]{Rieser2022}%
  \BibitemOpen
  \bibfield  {author} {\bibinfo {author} {\bibfnamefont {J.}~\bibnamefont
  {Rieser}}, \bibinfo {author} {\bibfnamefont {M.~A.}\ \bibnamefont
  {Ciampini}}, \bibinfo {author} {\bibfnamefont {H.}~\bibnamefont {Rudolph}},
  \bibinfo {author} {\bibfnamefont {N.}~\bibnamefont {Kiesel}}, \bibinfo
  {author} {\bibfnamefont {K.}~\bibnamefont {Hornberger}}, \bibinfo {author}
  {\bibfnamefont {B.~A.}\ \bibnamefont {Stickler}}, \bibinfo {author}
  {\bibfnamefont {M.}~\bibnamefont {Aspelmeyer}}, \ and\ \bibinfo {author}
  {\bibfnamefont {U.}~\bibnamefont {Delić}},\ }\href {\doibase
  10.48550/ARXIV.2203.04198} {\enquote {\bibinfo {title} {Observation of strong
  and tunable light-induced dipole-dipole interactions between optically
  levitated nanoparticles},}\ } (\bibinfo {year} {2022})\BibitemShut {NoStop}%
\bibitem [{\citenamefont {Tebbenjohanns}\ \emph
  {et~al.}(2019{\natexlab{b}})\citenamefont {Tebbenjohanns}, \citenamefont
  {Frimmer},\ and\ \citenamefont {Novotny}}]{Tebbenjohanns2019PRA}%
  \BibitemOpen
  \bibfield  {author} {\bibinfo {author} {\bibfnamefont {F.}~\bibnamefont
  {Tebbenjohanns}}, \bibinfo {author} {\bibfnamefont {M.}~\bibnamefont
  {Frimmer}}, \ and\ \bibinfo {author} {\bibfnamefont {L.}~\bibnamefont
  {Novotny}},\ }\href {\doibase 10.1103/PhysRevA.100.043821} {\bibfield
  {journal} {\bibinfo  {journal} {Phys. Rev. A}\ }\textbf {\bibinfo {volume}
  {100}},\ \bibinfo {pages} {043821} (\bibinfo {year}
  {2019}{\natexlab{b}})}\BibitemShut {NoStop}%
\bibitem [{\citenamefont {Hebestreit}\ \emph
  {et~al.}(2018{\natexlab{b}})\citenamefont {Hebestreit}, \citenamefont
  {Frimmer}, \citenamefont {Reimann}, \citenamefont {Dellago}, \citenamefont
  {Ricci},\ and\ \citenamefont {Novotny}}]{Hebestreit2018Calibration}%
  \BibitemOpen
  \bibfield  {author} {\bibinfo {author} {\bibfnamefont {E.}~\bibnamefont
  {Hebestreit}}, \bibinfo {author} {\bibfnamefont {M.}~\bibnamefont {Frimmer}},
  \bibinfo {author} {\bibfnamefont {R.}~\bibnamefont {Reimann}}, \bibinfo
  {author} {\bibfnamefont {C.}~\bibnamefont {Dellago}}, \bibinfo {author}
  {\bibfnamefont {F.}~\bibnamefont {Ricci}}, \ and\ \bibinfo {author}
  {\bibfnamefont {L.}~\bibnamefont {Novotny}},\ }\href {\doibase
  10.1063/1.5017119} {\bibfield  {journal} {\bibinfo  {journal} {Rev. Sci.
  Instrum.}\ }\textbf {\bibinfo {volume} {89}},\ \bibinfo {pages} {033111}
  (\bibinfo {year} {2018}{\natexlab{b}})}\BibitemShut {NoStop}%
\bibitem [{\citenamefont {Steixner}\ \emph {et~al.}(2005)\citenamefont
  {Steixner}, \citenamefont {Rabl},\ and\ \citenamefont
  {Zoller}}]{Steixner2005}%
  \BibitemOpen
  \bibfield  {author} {\bibinfo {author} {\bibfnamefont {V.}~\bibnamefont
  {Steixner}}, \bibinfo {author} {\bibfnamefont {P.}~\bibnamefont {Rabl}}, \
  and\ \bibinfo {author} {\bibfnamefont {P.}~\bibnamefont {Zoller}},\ }\href
  {\doibase 10.1103/PhysRevA.72.043826} {\bibfield  {journal} {\bibinfo
  {journal} {Phys. Rev. A}\ }\textbf {\bibinfo {volume} {72}},\ \bibinfo
  {pages} {043826} (\bibinfo {year} {2005})}\BibitemShut {NoStop}%
\bibitem [{\citenamefont {Bushev}\ \emph {et~al.}(2006)\citenamefont {Bushev},
  \citenamefont {Rotter}, \citenamefont {Wilson}, \citenamefont {Dubin},
  \citenamefont {Becher}, \citenamefont {Eschner}, \citenamefont {Blatt},
  \citenamefont {Steixner}, \citenamefont {Rabl},\ and\ \citenamefont
  {Zoller}}]{Bushev2006}%
  \BibitemOpen
  \bibfield  {author} {\bibinfo {author} {\bibfnamefont {P.}~\bibnamefont
  {Bushev}}, \bibinfo {author} {\bibfnamefont {D.}~\bibnamefont {Rotter}},
  \bibinfo {author} {\bibfnamefont {A.}~\bibnamefont {Wilson}}, \bibinfo
  {author} {\bibfnamefont {F.~m.~c.}\ \bibnamefont {Dubin}}, \bibinfo {author}
  {\bibfnamefont {C.}~\bibnamefont {Becher}}, \bibinfo {author} {\bibfnamefont
  {J.}~\bibnamefont {Eschner}}, \bibinfo {author} {\bibfnamefont
  {R.}~\bibnamefont {Blatt}}, \bibinfo {author} {\bibfnamefont
  {V.}~\bibnamefont {Steixner}}, \bibinfo {author} {\bibfnamefont
  {P.}~\bibnamefont {Rabl}}, \ and\ \bibinfo {author} {\bibfnamefont
  {P.}~\bibnamefont {Zoller}},\ }\href {\doibase 10.1103/PhysRevLett.96.043003}
  {\bibfield  {journal} {\bibinfo  {journal} {Phys. Rev. Lett.}\ }\textbf
  {\bibinfo {volume} {96}},\ \bibinfo {pages} {043003} (\bibinfo {year}
  {2006})}\BibitemShut {NoStop}%
\bibitem [{\citenamefont {Iwasaki}\ \emph {et~al.}(2019)\citenamefont
  {Iwasaki}, \citenamefont {Yotsuya}, \citenamefont {Naruki}, \citenamefont
  {Matsuda}, \citenamefont {Yoneda},\ and\ \citenamefont
  {Aikawa}}]{Iwasaki2018}%
  \BibitemOpen
  \bibfield  {author} {\bibinfo {author} {\bibfnamefont {M.}~\bibnamefont
  {Iwasaki}}, \bibinfo {author} {\bibfnamefont {T.}~\bibnamefont {Yotsuya}},
  \bibinfo {author} {\bibfnamefont {T.}~\bibnamefont {Naruki}}, \bibinfo
  {author} {\bibfnamefont {Y.}~\bibnamefont {Matsuda}}, \bibinfo {author}
  {\bibfnamefont {M.}~\bibnamefont {Yoneda}}, \ and\ \bibinfo {author}
  {\bibfnamefont {K.}~\bibnamefont {Aikawa}},\ }\href {\doibase
  10.1103/PhysRevA.99.051401} {\bibfield  {journal} {\bibinfo  {journal} {Phys.
  Rev. A}\ }\textbf {\bibinfo {volume} {99}},\ \bibinfo {pages} {051401}
  (\bibinfo {year} {2019})}\BibitemShut {NoStop}%
\bibitem [{\citenamefont {Cohadon}\ \emph {et~al.}(1999)\citenamefont
  {Cohadon}, \citenamefont {Heidmann},\ and\ \citenamefont
  {Pinard}}]{Cohadon1999}%
  \BibitemOpen
  \bibfield  {author} {\bibinfo {author} {\bibfnamefont {P.~F.}\ \bibnamefont
  {Cohadon}}, \bibinfo {author} {\bibfnamefont {A.}~\bibnamefont {Heidmann}}, \
  and\ \bibinfo {author} {\bibfnamefont {M.}~\bibnamefont {Pinard}},\ }\href
  {\doibase 10.1103/PhysRevLett.83.3174} {\bibfield  {journal} {\bibinfo
  {journal} {Phys. Rev. Lett.}\ }\textbf {\bibinfo {volume} {83}},\ \bibinfo
  {pages} {3174} (\bibinfo {year} {1999})}\BibitemShut {NoStop}%
\bibitem [{\citenamefont {Poggio}\ \emph {et~al.}(2007)\citenamefont {Poggio},
  \citenamefont {Degen}, \citenamefont {Mamin},\ and\ \citenamefont
  {Rugar}}]{Poggio2007}%
  \BibitemOpen
  \bibfield  {author} {\bibinfo {author} {\bibfnamefont {M.}~\bibnamefont
  {Poggio}}, \bibinfo {author} {\bibfnamefont {C.~L.}\ \bibnamefont {Degen}},
  \bibinfo {author} {\bibfnamefont {H.~J.}\ \bibnamefont {Mamin}}, \ and\
  \bibinfo {author} {\bibfnamefont {D.}~\bibnamefont {Rugar}},\ }\href
  {\doibase 10.1103/PhysRevLett.99.017201} {\bibfield  {journal} {\bibinfo
  {journal} {Phys. Rev. Lett.}\ }\textbf {\bibinfo {volume} {99}},\ \bibinfo
  {pages} {017201} (\bibinfo {year} {2007})}\BibitemShut {NoStop}%
\bibitem [{\citenamefont {Rossi}\ \emph {et~al.}(2018)\citenamefont {Rossi},
  \citenamefont {Mason}, \citenamefont {Chen}, \citenamefont {Tsaturyan},\ and\
  \citenamefont {Schliesser}}]{Rossi2018}%
  \BibitemOpen
  \bibfield  {author} {\bibinfo {author} {\bibfnamefont {M.}~\bibnamefont
  {Rossi}}, \bibinfo {author} {\bibfnamefont {D.}~\bibnamefont {Mason}},
  \bibinfo {author} {\bibfnamefont {J.}~\bibnamefont {Chen}}, \bibinfo {author}
  {\bibfnamefont {Y.}~\bibnamefont {Tsaturyan}}, \ and\ \bibinfo {author}
  {\bibfnamefont {A.}~\bibnamefont {Schliesser}},\ }\href {\doibase
  10.1038/s41586-018-0643-8} {\bibfield  {journal} {\bibinfo  {journal}
  {Nature}\ }\textbf {\bibinfo {volume} {563}},\ \bibinfo {pages} {53}
  (\bibinfo {year} {2018})}\BibitemShut {NoStop}%
\bibitem [{\citenamefont {Gieseler}\ \emph {et~al.}(2014)\citenamefont
  {Gieseler}, \citenamefont {Quidant}, \citenamefont {Dellago},\ and\
  \citenamefont {Novotny}}]{Gieseler2014}%
  \BibitemOpen
  \bibfield  {author} {\bibinfo {author} {\bibfnamefont {J.}~\bibnamefont
  {Gieseler}}, \bibinfo {author} {\bibfnamefont {R.}~\bibnamefont {Quidant}},
  \bibinfo {author} {\bibfnamefont {C.}~\bibnamefont {Dellago}}, \ and\
  \bibinfo {author} {\bibfnamefont {L.}~\bibnamefont {Novotny}},\ }\href
  {\doibase 10.1038/nnano.2014.40} {\bibfield  {journal} {\bibinfo  {journal}
  {Nature Nanotechnology}\ }\textbf {\bibinfo {volume} {9}},\ \bibinfo {pages}
  {358} (\bibinfo {year} {2014})}\BibitemShut {NoStop}%
\bibitem [{\citenamefont {Debnath}\ \emph {et~al.}(2016)\citenamefont
  {Debnath}, \citenamefont {Linke}, \citenamefont {Figgatt}, \citenamefont
  {Landsman}, \citenamefont {Wright},\ and\ \citenamefont
  {Monroe}}]{Debnath2016}%
  \BibitemOpen
  \bibfield  {author} {\bibinfo {author} {\bibfnamefont {S.}~\bibnamefont
  {Debnath}}, \bibinfo {author} {\bibfnamefont {N.~M.}\ \bibnamefont {Linke}},
  \bibinfo {author} {\bibfnamefont {C.}~\bibnamefont {Figgatt}}, \bibinfo
  {author} {\bibfnamefont {K.~A.}\ \bibnamefont {Landsman}}, \bibinfo {author}
  {\bibfnamefont {K.}~\bibnamefont {Wright}}, \ and\ \bibinfo {author}
  {\bibfnamefont {C.}~\bibnamefont {Monroe}},\ }\href {\doibase
  10.1038/nature18648} {\bibfield  {journal} {\bibinfo  {journal} {Nature}\
  }\textbf {\bibinfo {volume} {536}},\ \bibinfo {pages} {63} (\bibinfo {year}
  {2016})}\BibitemShut {NoStop}%
\bibitem [{\citenamefont {de~los R{\'i}os~Sommer}\ \emph
  {et~al.}(2021)\citenamefont {de~los R{\'i}os~Sommer}, \citenamefont {Meyer},\
  and\ \citenamefont {Quidant}}]{delosRiosSommer2021}%
  \BibitemOpen
  \bibfield  {author} {\bibinfo {author} {\bibfnamefont {A.}~\bibnamefont
  {de~los R{\'i}os~Sommer}}, \bibinfo {author} {\bibfnamefont {N.}~\bibnamefont
  {Meyer}}, \ and\ \bibinfo {author} {\bibfnamefont {R.}~\bibnamefont
  {Quidant}},\ }\href {\doibase 10.1038/s41467-020-20419-2} {\bibfield
  {journal} {\bibinfo  {journal} {Nature Communications}\ }\textbf {\bibinfo
  {volume} {12}},\ \bibinfo {pages} {276} (\bibinfo {year} {2021})}\BibitemShut
  {NoStop}%
\bibitem [{\citenamefont {Toro\ifmmode~\check{s}\else \v{s}\fi{}}\ \emph
  {et~al.}(2021)\citenamefont {Toro\ifmmode~\check{s}\else \v{s}\fi{}},
  \citenamefont {Deli\ifmmode~\acute{c}\else \'{c}\fi{}}, \citenamefont
  {Hales},\ and\ \citenamefont {Monteiro}}]{Toros2021}%
  \BibitemOpen
  \bibfield  {author} {\bibinfo {author} {\bibfnamefont {M.}~\bibnamefont
  {Toro\ifmmode~\check{s}\else \v{s}\fi{}}}, \bibinfo {author} {\bibfnamefont
  {U.~c.~v.}\ \bibnamefont {Deli\ifmmode~\acute{c}\else \'{c}\fi{}}}, \bibinfo
  {author} {\bibfnamefont {F.}~\bibnamefont {Hales}}, \ and\ \bibinfo {author}
  {\bibfnamefont {T.~S.}\ \bibnamefont {Monteiro}},\ }\href {\doibase
  10.1103/PhysRevResearch.3.023071} {\bibfield  {journal} {\bibinfo  {journal}
  {Phys. Rev. Research}\ }\textbf {\bibinfo {volume} {3}},\ \bibinfo {pages}
  {023071} (\bibinfo {year} {2021})}\BibitemShut {NoStop}%
\bibitem [{\citenamefont {Rudolph}\ \emph {et~al.}(2022)\citenamefont
  {Rudolph}, \citenamefont {Delić}, \citenamefont {Aspelmeyer}, \citenamefont
  {Hornberger},\ and\ \citenamefont {Stickler}}]{Rudolph2022}%
  \BibitemOpen
  \bibfield  {author} {\bibinfo {author} {\bibfnamefont {H.}~\bibnamefont
  {Rudolph}}, \bibinfo {author} {\bibfnamefont {U.}~\bibnamefont {Delić}},
  \bibinfo {author} {\bibfnamefont {M.}~\bibnamefont {Aspelmeyer}}, \bibinfo
  {author} {\bibfnamefont {K.}~\bibnamefont {Hornberger}}, \ and\ \bibinfo
  {author} {\bibfnamefont {B.~A.}\ \bibnamefont {Stickler}},\ }\href {\doibase
  10.48550/ARXIV.2204.13684} {\enquote {\bibinfo {title} {Force-gradient
  sensing and entanglement via feedback cooling of interacting
  nanoparticles},}\ } (\bibinfo {year} {2022})\BibitemShut {NoStop}%
\bibitem [{\citenamefont {Kamba}\ \emph {et~al.}(2022)\citenamefont {Kamba},
  \citenamefont {Shimizu},\ and\ \citenamefont {Aikawa}}]{Kamba2022}%
  \BibitemOpen
  \bibfield  {author} {\bibinfo {author} {\bibfnamefont {M.}~\bibnamefont
  {Kamba}}, \bibinfo {author} {\bibfnamefont {R.}~\bibnamefont {Shimizu}}, \
  and\ \bibinfo {author} {\bibfnamefont {K.}~\bibnamefont {Aikawa}},\ }\href
  {\doibase 10.48550/ARXIV.2205.00902} {\enquote {\bibinfo {title} {Optical
  cold damping of neutral nanoparticles near the ground state in an optical
  lattice},}\ } (\bibinfo {year} {2022})\BibitemShut {NoStop}%
\end{thebibliography}%


\begin{thebibliography}{4}%
\makeatletter
\providecommand \@ifxundefined [1]{%
 \@ifx{#1\undefined}
}%
\providecommand \@ifnum [1]{%
 \ifnum #1\expandafter \@firstoftwo
 \else \expandafter \@secondoftwo
 \fi
}%
\providecommand \@ifx [1]{%
 \ifx #1\expandafter \@firstoftwo
 \else \expandafter \@secondoftwo
 \fi
}%
\providecommand \natexlab [1]{#1}%
\providecommand \enquote  [1]{``#1''}%
\providecommand \bibnamefont  [1]{#1}%
\providecommand \bibfnamefont [1]{#1}%
\providecommand \citenamefont [1]{#1}%
\providecommand \href@noop [0]{\@secondoftwo}%
\providecommand \href [0]{\begingroup \@sanitize@url \@href}%
\providecommand \@href[1]{\@@startlink{#1}\@@href}%
\providecommand \@@href[1]{\endgroup#1\@@endlink}%
\providecommand \@sanitize@url [0]{\catcode `\\12\catcode `\$12\catcode
  `\&12\catcode `\#12\catcode `\^12\catcode `\_12\catcode `\%12\relax}%
\providecommand \@@startlink[1]{}%
\providecommand \@@endlink[0]{}%
\providecommand \url  [0]{\begingroup\@sanitize@url \@url }%
\providecommand \@url [1]{\endgroup\@href {#1}{\urlprefix }}%
\providecommand \urlprefix  [0]{URL }%
\providecommand \Eprint [0]{\href }%
\providecommand \doibase [0]{http://dx.doi.org/}%
\providecommand \selectlanguage [0]{\@gobble}%
\providecommand \bibinfo  [0]{\@secondoftwo}%
\providecommand \bibfield  [0]{\@secondoftwo}%
\providecommand \translation [1]{[#1]}%
\providecommand \BibitemOpen [0]{}%
\providecommand \bibitemStop [0]{}%
\providecommand \bibitemNoStop [0]{.\EOS\space}%
\providecommand \EOS [0]{\spacefactor3000\relax}%
\providecommand \BibitemShut  [1]{\csname bibitem#1\endcsname}%
\let\auto@bib@innerbib\@empty
\bibitem [{\citenamefont {Hebestreit}(2017)}]{Hebestreit2017}%
  \BibitemOpen
  \bibfield  {author} {\bibinfo {author} {\bibfnamefont {E.}~\bibnamefont
  {Hebestreit}},\ }\emph {\bibinfo {title} {Thermal Properties of Levitated
  Nanoparticles}},\ \href {\doibase 10.3929/ethz-b-000250832} {Ph.D. thesis},\
  \bibinfo  {school} {ETH Zurich}, \bibinfo {address} {Zurich} (\bibinfo {year}
  {2017})\BibitemShut {NoStop}%
\bibitem [{\citenamefont {Tebbenjohanns}\ \emph
  {et~al.}(2019{\natexlab{a}})\citenamefont {Tebbenjohanns}, \citenamefont
  {Frimmer},\ and\ \citenamefont {Novotny}}]{Tebbenjohanns2019PRA}%
  \BibitemOpen
  \bibfield  {author} {\bibinfo {author} {\bibfnamefont {F.}~\bibnamefont
  {Tebbenjohanns}}, \bibinfo {author} {\bibfnamefont {M.}~\bibnamefont
  {Frimmer}}, \ and\ \bibinfo {author} {\bibfnamefont {L.}~\bibnamefont
  {Novotny}},\ }\href {\doibase 10.1103/PhysRevA.100.043821} {\bibfield
  {journal} {\bibinfo  {journal} {Phys. Rev. A}\ }\textbf {\bibinfo {volume}
  {100}},\ \bibinfo {pages} {043821} (\bibinfo {year}
  {2019}{\natexlab{a}})}\BibitemShut {NoStop}%
\bibitem [{\citenamefont {Gieseler}\ \emph {et~al.}(2012)\citenamefont
  {Gieseler}, \citenamefont {Deutsch}, \citenamefont {Quidant},\ and\
  \citenamefont {Novotny}}]{Gieseler2012}%
  \BibitemOpen
  \bibfield  {author} {\bibinfo {author} {\bibfnamefont {J.}~\bibnamefont
  {Gieseler}}, \bibinfo {author} {\bibfnamefont {B.}~\bibnamefont {Deutsch}},
  \bibinfo {author} {\bibfnamefont {R.}~\bibnamefont {Quidant}}, \ and\
  \bibinfo {author} {\bibfnamefont {L.}~\bibnamefont {Novotny}},\ }\href
  {\doibase 10.1103/PhysRevLett.109.103603} {\bibfield  {journal} {\bibinfo
  {journal} {Phys. Rev. Lett.}\ }\textbf {\bibinfo {volume} {109}},\ \bibinfo
  {pages} {103603} (\bibinfo {year} {2012})}\BibitemShut {NoStop}%
\bibitem [{\citenamefont {Tebbenjohanns}\ \emph
  {et~al.}(2019{\natexlab{b}})\citenamefont {Tebbenjohanns}, \citenamefont
  {Frimmer}, \citenamefont {Militaru}, \citenamefont {Jain},\ and\
  \citenamefont {Novotny}}]{Tebbenjohanns2019colddamping}%
  \BibitemOpen
  \bibfield  {author} {\bibinfo {author} {\bibfnamefont {F.}~\bibnamefont
  {Tebbenjohanns}}, \bibinfo {author} {\bibfnamefont {M.}~\bibnamefont
  {Frimmer}}, \bibinfo {author} {\bibfnamefont {A.}~\bibnamefont {Militaru}},
  \bibinfo {author} {\bibfnamefont {V.}~\bibnamefont {Jain}}, \ and\ \bibinfo
  {author} {\bibfnamefont {L.}~\bibnamefont {Novotny}},\ }\href {\doibase
  10.1103/PhysRevLett.122.223601} {\bibfield  {journal} {\bibinfo  {journal}
  {Phys. Rev. Lett.}\ }\textbf {\bibinfo {volume} {122}},\ \bibinfo {pages}
  {223601} (\bibinfo {year} {2019}{\natexlab{b}})},\ \Eprint
  {http://arxiv.org/abs/1812.09875} {1812.09875} \BibitemShut {NoStop}%
\end{thebibliography}%

\end{document}


\scalefont{1.0}

\title{Supplementary Materials \\ Scalable all-optical cold damping of levitated nanoparticles}

\author{Jayadev~\surname{Vijayan}}
\email[Correspondence email address: ]{jvijayan@ethz.ch}
\author{Zhao~\surname{Zhang}}
\affiliation{Photonics Laboratory, ETH Z{\"u}rich, 8093 Z\"urich, Switzerland}
\author{Johannes~\surname{Piotrowski}}
\affiliation{Photonics Laboratory, ETH Z{\"u}rich, 8093 Z\"urich, Switzerland}
\author{Dominik~\surname{Windey}}
\affiliation{Photonics Laboratory, ETH Z{\"u}rich, 8093 Z\"urich, Switzerland}
\author{Fons~\surname{van der Laan}}
\affiliation{Photonics Laboratory, ETH Z{\"u}rich, 8093 Z\"urich, Switzerland}
\author{Martin~\surname{Frimmer}}
\affiliation{Photonics Laboratory, ETH Z{\"u}rich, 8093 Z\"urich, Switzerland}
\author{Lukas~\surname{Novotny}}
\affiliation{Photonics Laboratory, ETH Z{\"u}rich, 8093 Z\"urich, Switzerland}

\date{\today} 
\maketitle

\centerline{\textbf{Detection schemes}}
\label{sec:detection}
\vspace{1mm}

\begin{figure*}
    \centering
    \includegraphics[width = 17.2cm]{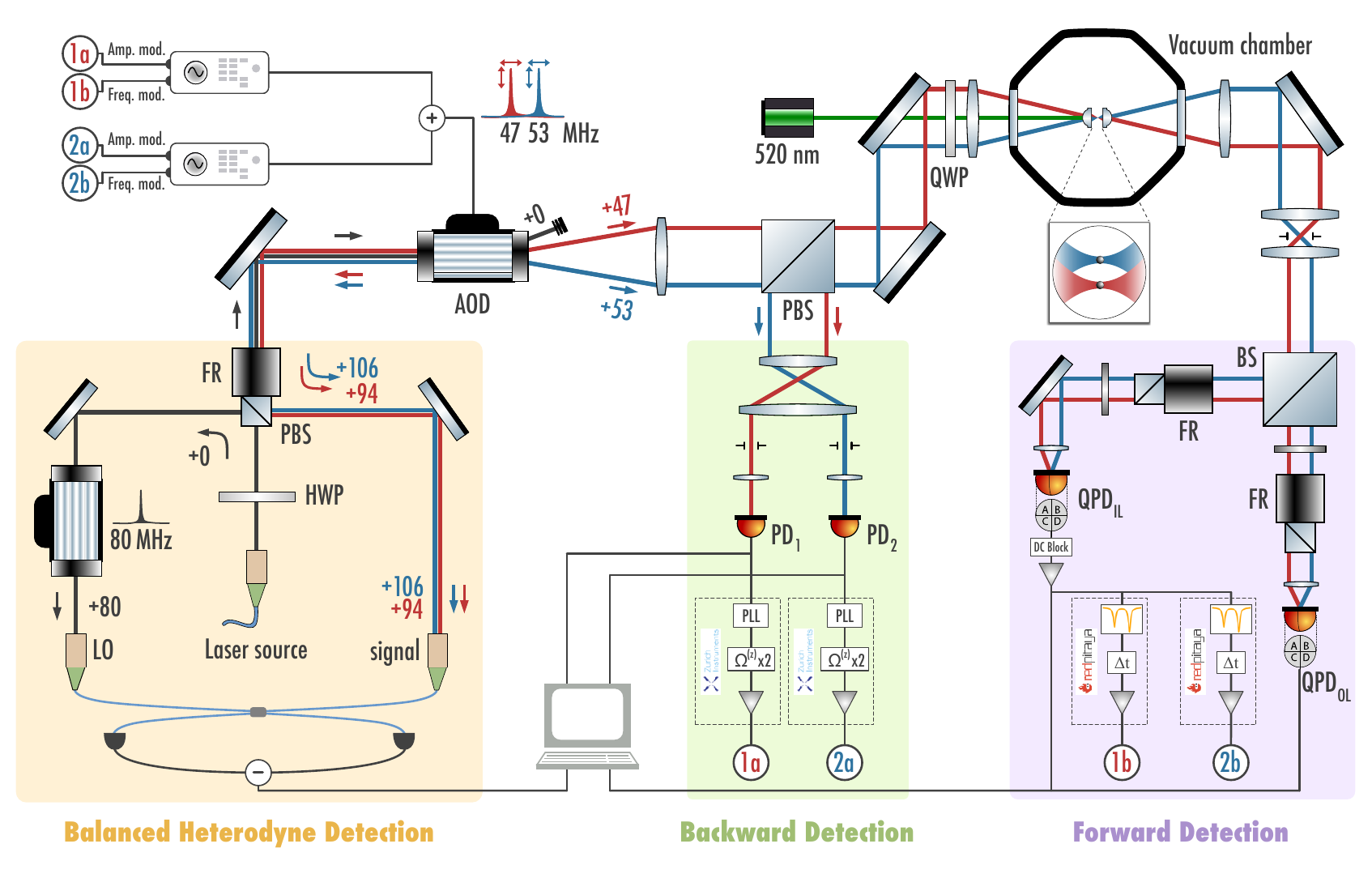}
    \caption{Figure S1. \textbf{Detection schemes used in the experiment.}
    Light scattered by the particle in the forward direction is detected on the QPD which performs measurement-based cold damping of the particle motion.
    The back-scattered light is detected on photodiodes that perform parametric cooling of the particle motion.
    Additionally, the back-scattered light is also used in a heterodyne detection scheme that can overcomes the scalability limitations of the forward detection.
    At the moment, it is used as a tool to detect multiple particles as they are loaded into the chamber.
    A green laser is used to illuminate the particles for taking high resolution images, such as in Fig. 4A of the main text.
    }
    \label{fig:S1}
\end{figure*}

A complete sketch of the optical setup is shown in Fig. S\ref{fig:S1}. 
As described in the main text, we have two quadrant photodetectors (QPDs) in the forward scattering direction.
One of the two QPDs, labelled $\text{QPD}_{\text{il}}$ serves as an in-loop detector for our optical linear feedback cooling.
The AC component of the voltage signal corresponding to the $y$-channel of $\text{QPD}_{\text{il}}$ is amplified and sent to a programmable FPGA (RedPitaya STEMlab 125-14) which carries out three functions: 1) uses notch frequency filters to suppress the frequency components not relevant for feedback cooling, 2) introduces a tunable time delay, which shifts the phase of the feedback signal and 3) amplifies the signal with a variable gain.
The signal is then sent as a frequency modulation input to the appropriate channel of the function generator (MOGLABS Agile RF Synthesizer) driving the acousto-optical deflector (AA Optoelectronics DSTX).
The second QPD in the forward direction, labelled $\text{QPD}_{\text{ol}}$ serves as an out-of-loop detector of particles motion.
It is a home-built QPD~\cite{Hebestreit2017} that is designed to withstand higher optical power ($50\,$mW) than the in-loop QPD (Thorlabs PDQ30C, $1\,$mW).

In the back-scattered direction, we use separate single-chip photodiodes, labelled $\text{PD}_{\text{i}}$ to measure the signal from either trapped particle.
These detectors are primarily sensitive to the $z$ motion of the particle~\cite{Tebbenjohanns2019PRA}.
The output signal is used to generate a new oscillator (Zurich Instruments MFLI), which is then doubled in frequency, phase shifted and amplified.
The output from the lock-in amplifier is fed back to the amplitude modulation input of the function generator and used to parametrically cool the $z$ motion of the particles.

Finally, we have a heterodyne detection scheme in the back-scattered direction capable of spectrally resolving the signal from any number of particles.
Typically, for two particle experiments, we add rf tones centered at $\omega_1 = 47\,$MHz and $\omega_2 = 53\,$MHz and send the output to the AOD to generate the two tweezers.
Consequently, the back-scattered light from the particle motion appear as sidebands around carriers at frequencies of $\omega_1 = 47\,$MHz and $\omega_2 = 53\,$MHz.
By appropriately tuning a quarter-wave plate before the chamber, the back-scattered light can be sent back towards the AOD rather that be split off to the photodiodes $PD_{\text{i}}$.
On passing through the AOD from the backwards direction, the tweezers are double-shifted in frequency (i.e, they now appear as sidebands around a carrier shifted by $2\times\omega_1 = 94\,$MHz and $2\times\omega_2 = 106\,$MHz).
The beams are spatially overlapped by construction, picked off using a combination of a Faraday rotator and a polarizing beam splitter, and coupled into an optical fiber.
The same beam splitter is used to split off a small fraction of the input light from the laser source and frequency shifted by $\omega_{\text{LO}} = 80\,$MHz to be used as a local oscillator, and coupled into a fiber.
The local oscillator at $\omega_{\text{LO}} = 80\,$MHz and the signal from the particles at $2\times\omega_1 = 94\,$MHz and $2\times\omega_2 = 106\,$MHz are then interfered and balanced to extract motional sidebands around carriers at $14\,$MHz and $26\,$MHz respectively.

\vspace{5mm}

\centerline{\textbf{Scalability of nanoparticle arrays}}
\label{sec:scale}
\vspace{1mm} 

One of the most significant differences between cold atoms and levitated nanoparticles is their mass.
Due to the relatively heavy mass of levitated nanoparticles (several fg), the optical power required to trap a single nanoparticle ($> 100\,$mW) is several orders of magnitude larger than for atoms.
Scaling up a levitated optomechanical system will naturally require using a larger amount of optical power.

Currently, we rely on the particles having different trap frequencies to perform independent feedback cooling.
The spectral separation between two motional peaks in the PSD needs to be at least $5\,$kHz for the notch filters to filter out the contribution from other particles.
Already with $200\,$mW per particle at the chamber, we are able to trap and spectrally resolve the signal from two particles.
The trap frequencies of particles scale with the square root of the optical power~\cite{Gieseler2012}.
By increasing the power to $1\,$W (commonly used optical power in single particle levitodynamics experiments), we can double our system size to $4$ particles and have sufficient spacing between motional frequencies.
A straightforward alternative solution to scale up the system would be to send the forward scattered light from each particle to a separate QPD for feedback.
In addition to scaling up the system, it would also allow us to independently control the dynamics of degenerate particles.

It is worth noting that the balanced heterodyne detection scheme does not depend on each trapped particle having a separate trap frequency.
Therefore, it can be scaled up to any number of particles, degenerate or not, as long as the double-shifted carriers are spectrally separated by at least $100\,$kHz, corresponding to the highest trap frequency of a single particle ($\Omega^{\text{(x)}} = 87\,$KHz in main text).
With the bandwidth of the AOD of $20\,$MHz, this corresponds to $200$ particles.
In our experiment, the balanced heterodyne scheme is currently fiber-based, leading to all motional information except the $z$ motion being lost.
Replacing the fibers with a free-space setup where the signal beams are interfered with an appropriate local oscillator directly on a QPD will allow us to feedback cool multiple particles with degenerate trap frequencies.

\vspace{5mm}
\centerline{\textbf{Linear feedback circuit for spatial modulation}}
\label{sec:linear}
\vspace{1mm}

\begin{figure}
    \includegraphics[width = 8.6cm]{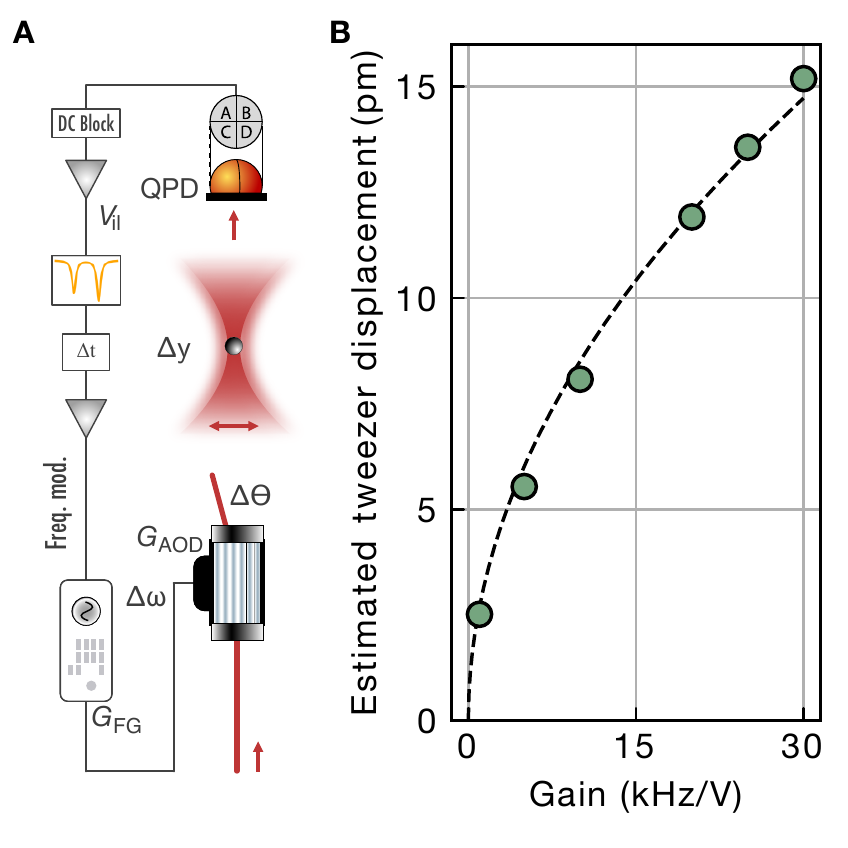}
    \caption{Figure S2. \textbf{Linear feedback circuit.}
    \textbf{A}.~A schematic of the cold damping feedback loop from the detected signal $V_{\text{il}}$ to the spatial displacement of the tweezer $\Delta y$.
    \textbf{B}.~Estimated tweezer displacement $\Delta y$ (green circles) for different gains applied at the function generator $G_{\text{FG}}$.
    The dashed line is a square root fit to the data (see text).
    }
    \label{fig:linearfeedback}
\end{figure}
The linear feedback loop used in the all-optical cold damping scheme is shown in detail in Fig. S\ref{fig:linearfeedback}A. 
The $y$ motion of the particle is detected by the in-loop QPD, whose output is DC-blocked, amplified and denoted as the in-looped signal $V_{\text{il}}$. 
It is then sent to an FPGA (RedPitaya) which first applies notch filters suppressing all frequencies that are not relevant.
The signal is then phase-shifted and amplified.
In our experiment, the gain factor at the RedPitaya is set to 1. 
The filtered signal is sent as frequency modulation input to the function generator driving the AOD. 
The resulting frequency modulation around the central frequency of the tone $\omega$ is $\Delta\omega=G_{\text{FG}}V_{\text{il}}$, where $G_{\text{FG}}$ is the controllable gain factor of frequency modulation at the function generator and ranges from 1kHz/V to 30kHz/V in the experiment. 
The frequency modulation of the rf signal driving the AOD leads to a modulating tilt angle of the diffracted first-order beam that eventually generates the trap: $\Delta \theta = G_{\text{AOD}}\Delta \omega$. 
Here, $G_{\text{AOD}}$ (in rad/Hz), the gain factor of AOD, is the ratio between its scan angle and input bandwidth. 
For the AOD we use, the factor is $G_{\text{AOD}} = 49\,$mrad$/20\,$MHz$=2.5\/\,$mrad/MHz.
The tilt in diffracted beam results in the spatial displacement of the tweezer position: $\Delta y=\Delta\theta\cdot f_{\text{eff}}/M$, where $M=0.6$ is the linear magnification of the telescope and $f_{\text{eff}}=0.5\,$mm is the effective focal length of the trapping lens inside the chamber. 
Finally, we obtain the total gain from the measured in-loop signal $V_{\text{il}}$ to the spatial modulation $\Delta y$ of the tweezer to be: $\Delta y=G V_{\text{il}} $ where $G\approx G_{\text{FG}}\cdot0.1$nm/kHz.
Since we are only interested in the c.m.\ motion of the particle along the $y$ direction, we apply a band-pass filter $F_y$ in the range $\Omega^{\text{(y)}} \pm 5\,$kHz.
Thus the amplitude of tweezer modulation during feedback cooling has the form: $\Delta y_{\text{fb}} = F_yG_{\text{FG}}V_{\text{il}}\cdot0.1\,$nm/kHz. 
The modulation amplitude as a function of feedback gain is shown in Fig.S\ref{fig:linearfeedback}B.

We can do a simple scaling analysis of the various quantities in the feedback loop as a function of the gain factor at rf function generator $G_{\text{FG}}$. 
The PSD contribution from the particle motion, which is proportional to the energy of the mode, is $A_{\text{particle}}\propto E=E_0\gamma_{\text{gas}}/\gamma_{\text{fb}} \propto 1/G_{\text{FG}}$, where $E$ ($E_0$) is the energy of the $y$ mode of a cooled (uncooled) particle. 
The in-loop voltage is proportional to the square root of the PSD: $V_{\text{il}} \propto \sqrt{A_{\text{particle}}} \propto 1/\sqrt{G_{\text{FG}}}$. 
The final tweezer modulation amplitude, which is the product of in-loop voltage and the total gain, is thus $\Delta y=GV_{\text{il}} \propto \sqrt{G_{\text{FG}}}$.
This matches the square root fitting of the tweezer motion in Fig. S\ref{fig:linearfeedback}B.

\vspace{5mm}

\centerline{\textbf{Reheating and ring-down measurements}}
\label{sec:reheating}
\vspace{1mm} 

\begin{figure}
    \includegraphics[width = 8.6cm]{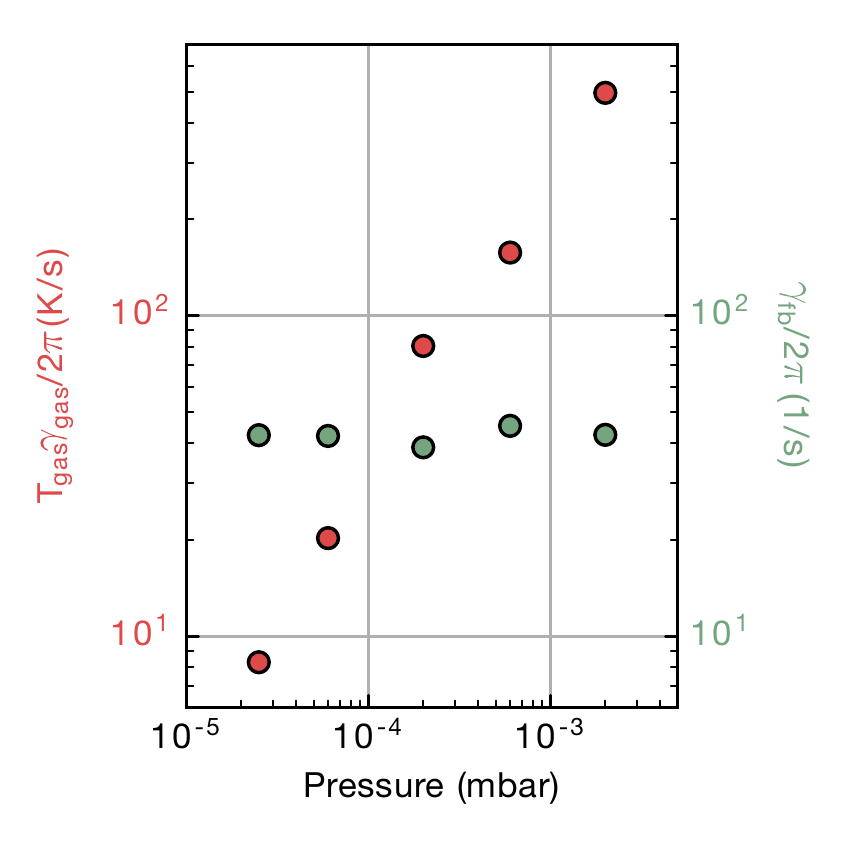}
    \caption{Figure S3. \textbf{Damping rates from ring-down and reheating measurements.} The feedback damping rate (green circles) is independent of pressure whereas the gas damping rate (red circles) increases with pressure.
    As in the main text, the gain is fixed to a low value of $\gamma_{\text{fb}} = 2\pi\times 42\,$Hz, corresponding to $G_{\text{FG}} = 5\,$kHz/V.
    }
    \label{fig:ringdown}
\end{figure}

\begin{figure}
    \includegraphics[width = 8.6cm]{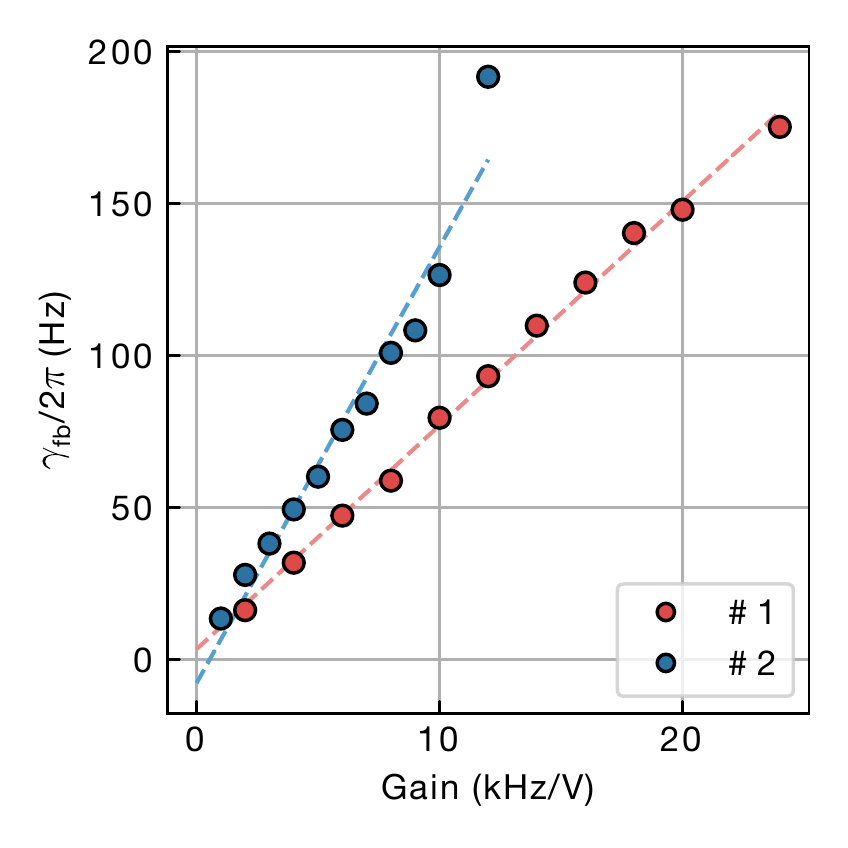}
    \caption{Figure S4. \textbf{Calibration of feedback gain.}
    Due to differences in the detection efficiency of the motional signal from particle 1 (red circles) and 2 (blue circles), the gain applied at the function generator $G_{\text{FG}}$ is adjusted to get the same $\gamma_\text{fb}$.
    Dashed lines are a linear fit to the data.
            }
    \label{fig:calib}
\end{figure}

With the non-equilibrium measurement protocol in the main text, we extract the feedback damping rate $\gamma_{\text{fb}}$ and the gas damping rate $T_{\text{gas}}\gamma_{\text{gas}}$ from ring-down and reheating cycles at fixed low function generator gain $G_{\text{FG}}=5\,$kHz/V over a  pressure range from $2\times 10^{-3}\,$mbar to $5\times 10^{-5}\,$mbar.
We show that across this pressure range, the c.m.\ temperatures measured from a calibrated PSD match the temperatures extracted from the damping rates, as predicted by Eq. 1 at low gain.
Tracking the non-equilibrium dynamics during the ring-down and reheating also allows us to directly measure $\gamma_{\text{fb}}$ and $T_{\text{gas}}\gamma_{\text{gas}}$ as a function of pressure.
We observe that the feedback damping rate at a fixed gain is independent of the pressure (see Fig.S\ref{fig:ringdown}), confirming that the damping rate under feedback is dominated by the cold damping feedback.
The gas damping rate, on the other hand, increases with gas pressure, as expected from previous experiments~\cite{Tebbenjohanns2019colddamping}.

The non-equilibrium experiments offer a way to reliably extract the damping rates, and consequently the temperature, independent of the effect of the tweezer motion. 
In the reheating process, the feedback is turned off and the gas damping rate $T_{\text{gas}}\gamma_{\text{gas}}$ is measured free from the tweezer motion. 
Although the feedback is turned on during the ring-down experiment, the extraction of the feedback damping rate $\gamma_{\text{fb}}$ is not affected by the tweezer modulation for the following reason. 
We assume that the energy of the particle motion during ring-down has the form $E(t)=E_0\exp(-\gamma_{\text{fb}t})$. 
Since the feedback circuit is linear, the total measured energy is $E'(t)=E_0(1+\eta) \exp(-\gamma_{\text{fb}t})$, where $\eta$ is the additional contribution from the tweezer motion, which, through feedback, is proportional to the contribution from particle motion. 
When we then perform an exponential fitting of the measured data $E'(t)$, we get the $\gamma_{\text{fb}}$ exactly the same as the damping rate of particle motion, regardless of the value of $\eta$.

\vspace{5mm}
\centerline{\textbf{Calibration of feedback gain}}
\label{sec:calib}
\vspace{1mm} 


As we alluded to in the main text, the detection efficiency of light scattered by each particle can be different, depending on experimental parameters and alignment.
Simultaneous cooling of different particles to the same temperature therefore requires adjusting the gain set in each feedback loop to result in the same feedback damping rate $\gamma_\text{fb}$.
We use the rf amplitude modulation control of the MOGLABs function generator to set the gain $G_{\text{FG}}$ in units of kHz/V.
To calibrate the feedback gain set at the function generator, we perform ringdown measurements for a wide range of gains and obtain the feedback cooling rate $\gamma_\text{fb}$ (Hz) as a function of applied gain $G_{\text{FG}}$ (kHz/V).
For the two particles used in Fig.~4 of the main text, the calibration is shown in Fig. S\ref{fig:calib}.
For the same feedback gain $G_{\text{FG}}$, the two particles experience a different feedback damping rates $\gamma_\text{fb}$.
In the experiment, for simultaneous cooling, we adjust the feedback gain settings to obtain the same feedback damping rates for both particles.

\bibliography{bibliography_supplementary}